\documentclass[a4paper,12pt]{article}

\usepackage{cite}
\usepackage{epsfig}
\usepackage{amssymb,amsmath}
\usepackage{mathrsfs}
\usepackage{bbm}
\usepackage{graphicx,subfigure}

\pdfoutput=1

\newlength{\dinwidth}
\newlength{\dinmargin}
\newcommand{\e}{\mathrm{e}}
\newcommand{\cR}{\mathcal{R}}

\usepackage{color}

\begin{document}

\title{
\begin{flushright}\vbox{\normalsize FTUV/12$-$0806 \\[-3pt] IFIC/12$-$27\\[-3pt] DO-TH 12/24}\end{flushright}\vskip 20pt
{\bf Exclusive radiative $B$-meson decays within\\[-5pt] the aligned two-Higgs-doublet model}}
\bigskip

\author{Martin Jung$^{1}$, Xin-Qiang Li$^{2,3}$ and Antonio Pich$^{3,4}$\\[20pt]
{$^1$\small Institut f\" ur Physik, Technische Universit\" at Dortmund, D-44221 Dortmund, Germany}\\
{$^2$\small Department of Physics, Henan Normal University, Xinxiang, Henan 453007, P.~R. China}\\
{$^3$\small IFIC, Universitat de Val\`encia-CSIC, Apt. Correus 22085, E-46071 Val\`encia, Spain}\\
 $^4$\small Physik Department and Institute for Advanced Study, \\[-10pt]
     \small Technische Universit\"at M\"unchen, D-85748 Garching, Germany}

\date{}
\maketitle
\bigskip \bigskip
\vspace{-1.cm}

\begin{abstract}
{\noindent}In the aligned two-Higgs-doublet model, the alignment of Yukawa matrices in flavour space guarantees the absence of tree-level flavour-changing neutral currents, while allowing at the same time for new sources of CP violation, implying potentially large effects in many low-energy processes. In this work we study the constraints from exclusive radiative $B\to V\gamma$ decays, where $V$ denotes a light vector meson. The current experimental data on the CP-averaged branching ratios and the direct CP and isospin asymmetries are analyzed. It is found that, while the branching ratios and direct CP asymmetries do not constrain the parameter space much further compared to the inclusive $B\to X_{s,d}\,\gamma$ decays, complementary constraints can be obtained from the isospin asymmetries $\Delta(K^*\gamma)$ and $\Delta(\rho\gamma)$. In addition, correlations between the various observables in exclusive $B\to V\gamma$ and inclusive $B\to X_{s,d}\,\gamma$ decays are investigated in detail, and predictions are made for several so far unmeasured observables.

\end{abstract}

\newpage

\section{Introduction}
\label{Sec:intro}

Within the Standard Model~(SM), flavour-changing neutral current~(FCNC) interactions, which are absent at tree level, are highly suppressed due to the Glashow-Iliopoulos-Maiani mechanism~\cite{Glashow:1970gm}, thereby providing a very sensitive probe of new physics~(NP). In this respect, the exclusive and inclusive radiative $B$-meson decays induced by the quark-level FCNC transitions $b \to s\gamma$ and $b \to d\gamma$ are of particular interest; for a recent review, see e.g. \cite{B2Vgreview}.

On the experimental side, especially the decays corresponding to $b\to s\gamma$ transitions are known with good accuracy, but branching ratios and even CP and isospin asymmetries have been measured for several $b\to d\gamma$ decays as well~\cite{Beringer:1900zz,Amhis:2012bh}. These results stem mainly from the two $B$-factory experiments Belle and BaBar; LHCb has however recently started to provide data as well~\cite{:2012qi,ICHEP:raddecays}. On the theoretical side, while the inclusive decays can be essentially calculated perturbatively with high precision, the exclusive processes are more complicated due to the involved interplay of non-perturbative strong interaction effects~\cite{B2Vgreview}. The QCD factorization~(QCDF) approach, which will be adopted in this paper, has provided a systematic framework for the treatment of exclusive radiative $B$-meson decays~\cite{Beneke:2001at,Bosch:2001gv,Ali:2001ez,Kagan:2001zk}. For alternative approaches, see e.g. \cite{B2Vg-PQCD,B2Vg-SCET,Ball:2006eu,Muheim:2008vu,B2Vg-LCSR}. Due to these experimental and theoretical improvements achieved in recent years, the exclusive $b \to s(d)\,\gamma$ decay channels are providing important constraints on various NP models~\cite{B2Vg-NP,B2Vg-NP-2}.

Due to its simplicity and being the low-energy limit of some more complete theories, the two-Higgs-doublet model~(2HDM) has been regarded as a popular extension of the SM since its first proposal in the 1970s~\cite{Lee:1973iz}. In its most general version, the model gives rise to unwanted FCNC phenomena due to the non-diagonal couplings of neutral scalars to fermions. Different ways to suppress them have been proposed, giving rise to a variety of specific implementations~\cite{2HDM-review}.

The tree-level FCNCs can be eliminated requiring the alignment in flavour space of the Yukawa matrices coupling to a given right-handed fermion~\cite{Pich:2010ic}. The aligned two-Higgs-doublet model~(A2HDM)~\cite{Pich:2009sp} results in a very specific structure, with all fermion-scalar interactions being proportional to the corresponding fermion masses. The only source of flavour-changing phenomena is the Cabibbo-Kobayashi-Maskawa~(CKM) quark mixing matrix~\cite{CKM}, appearing in the $W^\pm$ and charged scalar interactions. In addition to the fermion masses and the CKM matrix, the Yukawa Lagrangian is fully characterized in terms of three complex parameters $\varsigma_f$~($f=u, d, l$), which provide new sources for CP violation. The A2HDM leads to a rich and viable  phenomenology~\cite{Pich:2009sp,Jung:2010ik,Jung:2010ab}, with an interesting hierarchy of FCNC effects, suppressing them in light-quark systems while allowing potentially relevant signals in heavy-quark transitions.

In this paper, we study exclusive radiative $B$-meson decays within the A2HDM by employing the QCDF approach~\cite{Beneke:2001at,Bosch:2001gv}. In addition to branching ratios, we consider the CP and isospin asymmetries of these decays. Constraints on the relevant charged-scalar couplings to fermions are derived from the current data on these observables. Furthermore, we investigate correlations between these exclusive observables and the inclusive $B\to X_{s,d}\,\gamma$ decays, and predict several observables which have not been measured so far.

Our paper is organized as follows: In Sec.~\ref{Sec:B2Vgamma}, we first recapitulate the theoretical framework for $B\to V \gamma$ transitions and present the physical observables in these decays. In Sec.~\ref{Sec:A2HDM}, after briefly reviewing the A2HDM, we discuss its contributions to exclusive and inclusive $b\to s(d)\,\gamma$ decays. In Sec.~\ref{Sec:discussion}, we present and discuss our numerical results, including the correlations between exclusive and inclusive observables, before concluding in Sec.~\ref{Sec:conclusion}. The appendix includes a discussion of the relevant input parameters.

\section{$B\to V \gamma$ decays within the QCDF framework}
\label{Sec:B2Vgamma}

In this section, we briefly review the theoretical framework for $B\to V\gamma$ decays within the QCDF approach~\cite{Beneke:2001at,Bosch:2001gv}. For more details, the readers are referred to Refs.~\cite{Beneke:2001at,Bosch:2001gv,Ali:2001ez,Kagan:2001zk}.

\subsection{The weak effective Hamiltonian}
\label{Sec:Heff}

Within the SM, the weak effective Hamiltonian for radiative $b\to D\gamma$~($D=d, s$) transitions can be written as~\cite{Beneke:2001at}
\begin{equation}\label{Eq:Heff1}
 \mathcal H_{\rm eff} = -\frac{G_F}{\sqrt{2}}\left[\lambda_t^{(D)} {\cal H}_{\rm eff}^{(t)}
 +\lambda_u^{(D)} {\cal H}_{\rm eff}^{(u)} \right]\, + \mbox{h.c.},
\end{equation}
where $\lambda_q^{(D)}=V_{qb}V_{qD}^*$ are products of CKM matrix elements. For this result the unitarity relation $\lambda_u^{(D)}+\lambda_c^{(D)}+\lambda_t^{(D)}=0$ has been used; we have
\begin{eqnarray}\label{Eq:Heff2}
 \mathcal H_{\rm eff}^{(t)} &=& C_1\, \mathcal O_1^c + C_2\, \mathcal O_2^c +\sum_{i=3}^{8} C_i\, \mathcal O_i\,,\\
 \mathcal H_{\rm eff}^{(u)} &=& C_1\, (\mathcal O_1^c-\mathcal O_1^u) + C_2\, (\mathcal O_2^c-\mathcal O_2^u)\,.
\end{eqnarray}
For $b\to s$ transitions, since the term $\lambda_u^{(s)} {\cal H}_{\rm eff}^{(u)}$ is doubly Cabibbo-suppressed, its contribution to the decay amplitude is therefore very small.

We adopt the operator basis introduced by Chetyrkin, Misiak and M\"unz, which is characterized by a fully anticommuting $\gamma_5$ in dimensional regularization~\cite{Chetyrkin:1996vx},
\begin{align}\label{eq:operators}
   \mathcal O^p_1 &= \bar D \gamma_\mu (1-\gamma_5) T^a p\; \bar p \gamma^\mu (1-\gamma_5) T^a b\,,
  &\mathcal O^p_2 &= \bar D \gamma_\mu (1-\gamma_5) p\; \bar p \gamma^\mu (1-\gamma_5) b\,, \nonumber \\[0.2cm]
   \mathcal O_3 &= 2\, \bar D \gamma_\mu (1-\gamma_5) b\; \sum_q\, \bar q\gamma^\mu q\,,
  &\mathcal O_5 &= 2\, \bar D \gamma_{\mu_1}\gamma_{\mu_2}\gamma_{\mu_3} (1-\gamma_5) b\; \sum_q \,\bar q
   \gamma^{\mu_1}\gamma^{\mu_2}\gamma^{\mu_3} q\,, \nonumber \\
   \mathcal O_4 &= 2\, \bar D \gamma_\mu (1-\gamma_5) T^a b\; \sum_q\, \bar q\gamma^\mu T^a q\,,
  &\mathcal O_6 &= 2\, \bar D \gamma_{\mu_1}\gamma_{\mu_2}\gamma_{\mu_3} (1-\gamma_5) T^a b\; \sum_q \, \bar q
   \gamma^{\mu_1}\gamma^{\mu_2}\gamma^{\mu_3} T^a q\,, \nonumber \\
   \mathcal O_7 &= -\frac{e}{8\pi^2}\, \overline{m}_b\, \bar D\sigma^{\mu\nu} (1+\gamma_5) b\, F_{\mu\nu}\,,
  &\mathcal O_8 &= -\frac{g_s}{8\pi^2}\, \overline{m}_b\, \bar D\sigma^{\mu\nu}T^a (1+\gamma_5) b\, G^a_{\mu\nu}\,,
\end{align}
where $T^a~(a=1,\dots,8)$ stands for the $\mathrm{SU(3)_C}$ generators, $\overline m_b$ denotes the $b$-quark mass in the $\overline{\rm MS}$ scheme, and $e~(g_s)$ is the electromagnetic~(strong) coupling constant. The corresponding Wilson coefficients can be calculated using renormalization-group-improved perturbation theory~\cite{Chetyrkin:1996vx,WC1,WC2}. For convenience, we collect in Table~\ref{tab:WC} their numerical values at the scale $\mu=4.4~{\rm GeV}$ in leading-logarithmic~(LL) and next-to-leading logarithmic~(NLL) approximation, where $C_7^{\rm eff}=C_7-C_3/3-4 C_4/9-20 C_5/3-80 C_6/9$ and $C_{8}^{\rm eff}=C_8+C_3- C_4/6+20 C_5-10 C_6/3$ are the so-called ``effective coefficients"~\cite{Chetyrkin:1996vx}.

\begin{table}[thb]
\begin{center}
\caption{\label{tab:WC} \small Wilson coefficients at the scale $\mu=4.4~{\rm GeV}$ in LL and NLL approximation, using two-loop running for $\alpha_s$ and the input parameters listed in the appendix.}
\vspace{0.2cm}
\doublerulesep 0.8pt \tabcolsep 0.11in
\begin{tabular}{lcccccccc}
\hline\hline
    & ${C}_1$   & ${C}_2$  & ${C}_3$   & ${C}_4$   & ${C}_5$  & ${C}_6$  & $C_7^{\rm eff}$ & $C_8^{\rm eff}$ \\
\hline
LL  & $-0.5166$ & $1.0263$ & $-0.0052$ & $-0.0698$ & $0.0005$ & $0.0011$ & $-0.3126$ & $-0.1488$ \\
NLL & $-0.3062$ & $1.0083$ & $-0.0048$ & $-0.0840$ & $0.0003$ & $0.0009$ & $-0.3062$ & $-0.1682$ \\
\hline\hline
\end{tabular}
\end{center}
\end{table}

\subsection{Factorization formula for the hadronic matrix elements}
\label{Sec:matrixelement}

Starting from the weak effective Hamiltonian in Eq.~(\ref{Eq:Heff1}), the hadronic matrix elements for exclusive $B\to V\gamma$ decays can be written in the heavy $b$-quark limit  as~\cite{Beneke:2001at,Bosch:2001gv}
\begin{equation}\label{eq:matrixelement}
 \langle V(p^{\prime},\varepsilon) \gamma(q,\eta) | {\cal H}^{(i)}_{\rm eff} | \overline{B}(p) \rangle
 = \frac{i e m_b}{2\pi^2}\, {\cal T}_\perp^{(i)}(0)\, \bigg\{\epsilon^{\mu\nu\rho\sigma}\, \eta^{\ast}_{\mu}\,
 \varepsilon^{\ast}_{\nu}\, p_{\rho} p^{\prime}_{\sigma} - i\, \Big[(\eta^{\ast} \cdot \varepsilon^{\ast})\,
 (p^{\prime} \cdot q) - (\eta^{\ast} \cdot p^{\prime})\, (\varepsilon^{\ast} \cdot q)\Big]\bigg\},
\end{equation}
where the kinematic variables of initial and final states are indicated in the parentheses, and the Bjorken-Drell convention for the Levi-Civita tensor, $\epsilon^{0123}=-1$, is adopted. The light vector-meson state $|V\rangle$ is defined as
\begin{equation}
 |V\rangle \;\equiv\; \left\{\begin{array}{l}
          |\rho^-\rangle\,, |K^{*-} \rangle\, \qquad\qquad\qquad \mbox{for $B^-$-meson decays}\,, \\[0.2cm]
 -\sqrt 2 |\rho^0\rangle\,, \sqrt 2 |\omega\rangle\,, |\overline{K}^{*0}\rangle \quad\, \mbox{for $\overline{B}^0$-meson decays}\,, \\[0.2cm]
          |\phi \rangle\,, |K^{*0}\rangle\, \qquad\,\qquad\,\quad\,\quad\, \mbox{for $\overline{B}_s$-meson decays}\,.
\end{array}\right.
\end{equation}
With the above convenient parametrization, all the dynamical information is encoded in the function ${\cal T}_\perp^{(i)}(0)$, the calculation of which constitutes a big challenge~\cite{Beneke:2001at,Bosch:2001gv,Ali:2001ez}.

In the QCDF formalism, the function ${\cal T}_\perp^{(i)}(0)$ can be computed at leading power in a $\Lambda_{\rm QCD}/m_b$ expansion in terms of $B\to V$ transition form factors and light-cone distribution amplitudes~(LCDAs). Explicitly, the following factorization formula holds~\cite{Beneke:2001at}:
\begin{equation}\label{calT}
 {\cal T}_\perp^{(i)}(0)\; =\; T_1(0) \, C_\perp^{(i)}\,  +\, \frac{\pi^2}{N_c}\,\frac{f_B f_V^\perp}{m_B}\, \sum_{\pm}
 \int\frac{d\omega}{\omega}\, \Phi_{B,\,\pm}(\omega) \int_0^1 \!du\,\, \phi_\perp(u) \, T_{\perp,\,\pm}^{(i)}(u,\omega)\,,
\end{equation}
where $f_B$ and $\Phi_{B,\pm}(\omega)$ denote the $B$-meson decay constant and LCDAs, and $f_V^\perp$ and $\phi_{\perp}(u)$ the corresponding quantities of the transversely polarized vector meson. The first term is expressed in terms of the tensor form factor $T_1(0)$, and corresponds to vertex contributions, where the spectator quark in the $B$ meson does not participate in the hard process. The second term, on the other hand, incorporates the hard scattering of the spectator quark. Accordingly, they are referred to as the ``form factor" and the ``spectator scattering" term, respectively.

The hard-scattering kernels $C_\perp^{(i)}$ and $T_{\perp,\,\pm}^{(i)}(u,\omega)$ in Eq.~(\ref{calT}) are perturbatively calculable within the QCDF framework. Up to next-to-leading order~(NLO) in $\alpha_s$, they have the following expansions~\cite{Beneke:2001at}:
\begin{align}
 C_\perp^{(i)}\; &=\; C_\perp^{(0,i)}+\frac{\alpha_s C_F}{4\pi} \,C_\perp^{(1,i)} + \ldots, \label{c11} \\[0.2cm]
 T^{(i)}_{\perp,\,\pm}(u,\omega)\; &=\; T^{(0,i)}_{\perp,\,\pm}(u,\omega) + \frac{\alpha_s C_F}{4\pi} \,
 T^{(1,i)}_{\perp,\,\pm}(u,\omega) + \ldots\,, \label{t11}
\end{align}
where the strong coupling $\alpha_s$ should be evaluated at the scale $\mu_b \simeq m_b$ in Eq.~(\ref{c11}) and at $\mu_{h} \simeq (m_b\Lambda_{\rm QCD})^{1/2}$ in Eq.~(\ref{t11}), corresponding to the typical virtualities in these two terms. Details about the calculation and explicit expressions for the coefficients $C_\perp^{(0,i)}$, $C_\perp^{(1,i)}$ and $T^{(0,i)}_{\perp,\,\pm}(u,\omega)$, $T^{(1,i)}_{\perp,\,\pm}(u,\omega)$ can be found in Ref.~\cite{Beneke:2001at}.

In addition to the leading-power contributions given by Eqs.~(\ref{c11}) and (\ref{t11}), it is also well-known that specific power corrections, such as the weak annihilation in $B \to \rho\gamma$ decays, are numerically important~\cite{Beneke:2001at,Bosch:2001gv,Ali:2001ez}. Most importantly, the annihilation topologies have been shown to provide the main source of isospin asymmetry in $B \to K^* \gamma$ decays~\cite{Kagan:2001zk}. Despite being power-suppressed in $\Lambda_{\rm QCD}/m_b$, these terms are still computable within the QCDF framework. Thus, they are also included in this paper, and denoted by $\Delta\mathcal T_\perp^{(i)}\vert_{\rm ann}$ and $\Delta\mathcal T_\perp^{(i)}\vert_{\rm hsa}$. Explicit expressions for these terms can be found in Refs.~\cite{Beneke:2001at,Feldmann:2002iw}.

However, it should be noted that an endpoint divergence is encountered in the matrix element of the chromo-magnetic dipole operator $\mathcal O_8$, belonging to the term $\Delta\mathcal T_\perp^{(t)}\vert_{\rm hsa}$. This is a well-analyzed problem for exclusive $B$-meson decays within the framework of QCDF and/or its field-theoretical formulation, the soft-collinear effective theory~(SCET)~\cite{SCET1,SCET2}. Unfortunately, there is currently no satisfactory solution to this problem within the QCDF/SCET methods~\cite{QCDF-FF,Becher:2003qh,Arnesen:2006vb}. Following the treatment adopted in Refs.~\cite{Kagan:2001zk,Feldmann:2002iw}, we regulate this singularity with an \textit{ad-hoc} cutoff
\begin{equation}
  \int_0^1 du \;\rightarrow\; (1+\rho \, e^{i \chi}) \, \int_0^{1-\Lambda_h/m_B} du\,,
\end{equation}
and take $\Lambda_h \simeq 0.5~{\rm GeV}$, together with $0 \leq \rho\leq 1$ and $0 \leq \chi \leq 2 \pi$, to give an estimate of the theoretical uncertainty related to this power correction.

\subsection{Observables in $B\to V\gamma$ decays}
\label{Sec:observables}

In order to define the physical observables in $B\to V\gamma$ decays, it is more convenient to express the decay amplitude in terms of a new quantity ${\cal C}_7^{(i)}$, which is defined by~\cite{Beneke:2001at}
\begin{equation} \label{CalC7}
 {\cal C}_7^{(i)} \;\equiv\; \frac{{\cal T}_{\perp}^{(i)}(0)}{T_1(0)}\; =\; \delta^{it}\,C_7^{\rm eff} + \ldots\,,
\end{equation}
where $i=t, u$ refers to the two different CKM factors, and the ellipses denote the subleading perturbative and power corrections discussed in the previous subsection.

In terms of the quantity ${\cal C}_7^{(i)}$, the decay rate for $\overline{B}\to V\gamma$ decays can be written as~\cite{Beneke:2001at}
\begin{equation}\label{Gamma}
  \Gamma(\overline{B}\to V \gamma)\; =\; \frac{G_F^2}{8\pi^3}\,m_B^3\, S\left(1-\frac{m_V^2}{m_B^2}\right)^{\!3}
  \frac{\alpha_{\rm em}}{4\pi}\,m_{b}^2\,T_1(0)^2\, \left|\lambda_t^{(D)}{\cal C}_7^{(t)} + \lambda_u^{(D)}{\cal C}_7^{(u)}\right|^2\,,
\end{equation}
where $G_F$ is the Fermi coupling constant, $\alpha_{\rm em}=e^2/4\pi$ the fine-structure constant, and $S=1/2$ for $\rho^0$ and $\omega$ mesons, while $S=1$ for the other light vector mesons. Within the SM, the decay rate for the CP-conjugate mode, $\Gamma(B\to \overline{V} \gamma)$, can be obtained from Eq.~(\ref{Gamma}) with the replacement $\lambda_i^{(D)}\to \lambda_i^{(D)\ast}$. For $b\to s$ transitions, the dominant contribution comes from $\lambda_t^{(s)}{\cal C}_7^{(t)}$, since the term proportional to $\lambda_u^{(s)}$ is doubly Cabibbo suppressed, whereas for $b\to d$ transitions $|\lambda_u^{(d)}|$ is of the same order as $|\lambda_t^{(d)}|$, and the interference between them is the main source of CP-violating and isospin-breaking effects.

With the decay rate in Eq.~(\ref{Gamma}) at hand, the interesting observables in $B\to V\gamma$ decays can be defined as follows~\cite{Beneke:2001at,Bosch:2001gv,Ali:2001ez}: the CP-averaged branching ratio
\begin{equation}\label{Br}
 \mathrm{Br}(B\to V \gamma)\;  = \tau_B\,\bar{\Gamma}(B\to V\gamma)=\; \tau_B\; \frac{\Gamma(\overline{B}\to V \gamma) + \Gamma(B\to \overline{V} \gamma)}{2}\,,
\end{equation}
with $\tau_B$ denoting the $B$-meson lifetime, the direct CP asymmetry
\begin{equation}\label{ACP}
 A_{CP}(B\to V\gamma)\; =\; \frac{\Gamma(\overline{B} \to V \gamma) - \Gamma(B \to \overline{V} \gamma)}{
 \Gamma(\overline{B} \to V \gamma) + \Gamma(B \to \overline{V} \gamma)}\,,
\end{equation}
and the isospin asymmetries
\begin{align}\label{Isospin}
 \Delta(K^*\gamma)\; &=\;\frac{\bar\Gamma(B^0\to K^{*0}\gamma)-\bar\Gamma(B^+\to K^{*+}\gamma)}{\bar\Gamma(B^0\to
 K^{*0}\gamma)+\bar\Gamma(B^+\to K^{*+}\gamma)}, \\[0.2cm]
 \Delta(\rho\gamma)\; &=\;\frac{\bar\Gamma(B^+\to\rho^+\gamma)}{2\bar\Gamma(B^0\to\rho^0\gamma)}-1\,.
\end{align}
These observables can be used to test the SM and to probe various NP scenarios. Especially the two isospin asymmetries $\Delta(K^*\gamma)$ and $\Delta(\rho\gamma)$ provide complementary information to the  corresponding inclusive decay modes~\cite{B2Vg-NP,B2Vg-NP-2}.

Note that we do not discuss the indirect CP violation. The reason is that this observable remains proportional to $m_D/m_b$ for a $b\to D$ transition, rendering it very small in the decays considered here. Observation of a significant non-zero value would therefore imply NP beyond the A2HDM. We note, however, that the available measurements are compatible with zero~\cite{Amhis:2012bh}.

\section{The aligned two-Higgs-doublet model}
\label{Sec:A2HDM}

\subsection{Overview of the A2HDM}
\label{Sec:A2HDM-overview}

The 2HDM extends the SM with a second scalar doublet of hypercharge $Y=\frac{1}{2}$. Thus, in addition to the three needed Goldstone bosons, it contains five physical scalars: two charged fields $H^{\pm}$ and three neutral ones $\varphi_i^0=\{h,H,A\}$. The neutral mass-eigenstates $\varphi_i^0$ are related to the original scalar-doublet neutral fields $S_i$ through an orthogonal transformation $\varphi_i^0=\mathcal{R}_{ij}S_j$.

The most generic Yukawa Lagrangian with the SM fermionic content gives rise to FCNCs because the Yukawa couplings of the two scalar doublets to fermions cannot be simultaneously diagonalized in flavour space. The non-diagonal neutral couplings can be eliminated by requiring the alignment in flavour space of the Yukawa matrices~\cite{Pich:2009sp}; i.e., the two Yukawa matrices coupling to a given type of right-handed fermions are assumed to be proportional to each other and can, therefore, be diagonalized simultaneously. The three proportionality parameters $\varsigma_f$~($f=u,d,l$) are arbitrary complex numbers and introduce new sources of CP violation.

In terms of the fermion mass-eigenstate fields, the Yukawa interactions of the A2HDM read~\cite{Pich:2009sp}
\begin{equation}\label{lagrangian}
 \mathcal L_Y =  - \frac{\sqrt{2}}{v}\, H^+ \left\{ \bar{u} \left[ \varsigma_d\, V M_d \mathcal P_R - \varsigma_u\, M_u^\dagger V \mathcal P_L \right]  d\, + \, \varsigma_l\, \bar{\nu} M_l \mathcal P_R l \right\}\,-\,\frac{1}{v}\, \sum_{\varphi, f}\, y^{\varphi^0_i}_f\, \varphi^0_i\, \bar{f}\, M_f \mathcal P_R f\,+\,\mathrm{h.c.}
\end{equation}
where $V$ denotes the CKM matrix, $\mathcal P_{R,L}\equiv \frac{1\pm \gamma_5}{2}$ are the chirality projectors, and the neutral scalar couplings are given by
\begin{equation}
y_{d,l}^{\varphi^0_i} = \cR_{i1} + (\cR_{i2} + i\,\cR_{i3})\,\varsigma_{d,l}\,,
\qquad\qquad
y_u^{\varphi^0_i} = \cR_{i1} + (\cR_{i2} -i\,\cR_{i3}) \,\varsigma_{u}^*\,.
\end{equation}

From Eq.~(\ref{lagrangian}) we can see that, in the A2HDM, all fermionic couplings to scalars are proportional to the corresponding fermion masses, and the neutral-current interactions are diagonal in flavour. The only source of flavour-changing interactions is the CKM mixing matrix in the charged-current quark sector. All possible freedom allowed by the alignment conditions is determined by the three family-universal complex parameters $\varsigma_f$, which provide new sources of CP violation without tree-level FCNCs~\cite{Pich:2009sp}.\footnote{The alignment parameters $\varsigma_f$ are invariant under global SU(2) transformations of the two scalar doublets, $\phi_a \to \phi'_a = U_{ab} \,\phi_b$; {\it i.e.}, they are independent of the basis choice adopted in the scalar space. Given an arbitrary basis, where the two scalar doublets have vacuum expectation values $\langle \phi^T_a\rangle = \frac{1}{\sqrt{2}}\, (0, v_a \,\mathrm{e}^{i\theta_a})$~($a=1,2$), $\varsigma_f$ is a function of $v_2/v_1$, $\theta_2-\theta_1$ and the (complex) proportionality parameter between the two aligned Yukawa matrices coupling to the right-handed fermion $f_R$~\cite{Pich:2009sp}. In the so-called `Higgs basis'~\cite{2HDM-review}, where only one scalar doublet acquires a vacuum expectation value, $\varsigma_f$ is just the proportionality parameter between the two aligned Yukawas coupling to the right-handed fermion $f_R$~\cite{Pich:2009sp}.}

Higher-order corrections induce a misalignment of the Yukawa matrices, generating small FCNC effects suppressed by the corresponding loop factors. However, the flavour symmetries of the A2HDM tightly constraint the possible FCNC structures, keeping their effects well below the present experimental bounds~\cite{Pich:2010ic,Pich:2009sp,Jung:2010ik,Yukawa-alignment}. Although all possible sources of CP violation~(from the Yukawa matrices as well as the scalar potential) are taken into account in the A2HDM~\cite{Pich:2009sp}, to the order we are working, the radiative $B$-meson decays are only sensitive to charged scalar exchange; therefore, the relevant CP-violating effects originate in the parameters $\varsigma_f$ and the CKM phase. Thus, in this paper we shall focus only on the phenomenology of the charged-scalar Yukawa Lagrangian given by the first term in Eq.~(\ref{lagrangian}).

The presence of flavour-blind phases could induce electric dipole moments~(EDMs) at a measurable level~\cite{Buras:2010zm}. Direct one-loop contributions to the light-quark EDMs are strongly suppressed by the light quark masses and/or CKM factors. However, this suppression is no longer present in some two-loop contributions involving scalar exchanges and a heavy-quark loop~\cite{EDMs-1,EDMs-2}. For values of $|\mathrm{Im}(\varsigma_u^*\varsigma_d)|\lesssim 1$, the predicted neutron and mercury EDMs are smaller than the present experimental upper bounds, but they could be within the reach of future high-precision measurements. A detailed analysis of EDMs within the A2HDM is in progress~\cite{edm-a2hdm}.

\subsection{The A2HDM effects in exclusive $B\to V \gamma$ decays}
\label{Sec:B2VgammawithinA2HDM}

In a 2HDM without tree-level FCNCs, the NP contribution to $b\to s(d)\, \gamma$ transitions comes only from the charged-scalar penguin diagrams. In the approximation of vanishing strange quark mass, the resulting effective low-energy operator basis remains the same as in the SM, and the charged-scalar effect appears only in the short-distance Wilson coefficients at the matching scale $\mu_W$. The charged-scalar contribution to the matching condition has been calculated up to NLO independently by several groups~\cite{Borzumati:1998tg,Ciuchini:1997xe,Ciafaloni:1997un,Bobeth:1999ww,Degrassi:2010ne}.

Specific to the A2HDM and up to NLO, it is found that the nonzero charged-scalar contribution to the Wilson coefficients at the matching scale $\mu_W$ resides only in $C_{4,7,8}^{\mathrm{eff}}(\mu_W)$. For convenience, we can render explicit their dependence on the couplings $\varsigma_{u,d}$, with the following compact form:
\begin{equation}\label{eq::Cstructure}
 C_i^{\mathrm{eff}}(\mu_W) = C_{i,\mathrm{SM}}^{\mathrm{eff}}(\mu_W) + |\varsigma_u|^2\,C_{i,uu} - \varsigma_u^*\varsigma_d\,C_{i,ud}\,,
\end{equation}
where $\varsigma_u^*\varsigma_d=|\varsigma_u||\varsigma_d|\, \e^{-i\phi}$. The short-distance coefficients $C_{i,uu}$ and $C_{i,ud}$ are dominated by the top-quark contributions, and their explicit expressions at the LO and NLO can be extracted from Refs.~\cite{Borzumati:1998tg,Ciuchini:1997xe,Ciafaloni:1997un,Bobeth:1999ww,Degrassi:2010ne}.

From Eq.~(\ref{eq::Cstructure}), it can be seen that, depending on the relative phase $\phi$, the combined effect of the two terms $C_{i,uu}$ and $C_{i,ud}$ can be rather different. In the calculation of these matching coefficients, all the heavy particles~(including the top quark, vector bosons and the charged Higgs boson) have been integrated out simultaneously at the scale $\mu_W$; this is a reasonable approximation provided that the charged-scalar mass $m_{H^{\pm}}$ is of the same order of magnitude as $m_{W}$ and $m_t$. The evolution of the Wilson coefficients from the matching scale $\mu_W$ down to the low-energy scale $\mu_b\simeq m_b$ remains the same as in the SM. Details about the renormalization group evolution and the corresponding solution can be found e.g. in Refs.~\cite{Chetyrkin:1996vx,WC1,WC2}.

\section{Numerical results and discussion}
\label{Sec:discussion}

With the theoretical framework presented in the previous sections and the input parameters collected in the appendix, we are prepared to present and discuss our numerical results in this section.

\subsection{SM predictions and experimental data}
\label{Sec:smanddata}

\begin{table}[thb]
\begin{center}
\caption{\label{tab:SMPredictions} \small SM predictions for exclusive $B\to V\gamma$ decays within the QCDF framework. The branching ratios are given in units of $10^{-6}$, the direct CP and isospin asymmetries in units of $10^{-2}$. For completeness, our predictions for inclusive $B\to X_s \gamma$ and $B\to X_d \gamma$ decays are presented as well$^\dagger$. The theoretical uncertainties are combined by adding them in quadrature.}
\vspace{0.2cm}
\doublerulesep 0.8pt \tabcolsep 0.50in
\begin{tabular}{lcc}
\hline \hline
Observable & Exp. data~\cite{Amhis:2012bh} & SM prediction \\
\hline
  $\mathrm{Br}(B^+ \to K^{*+}\gamma)$ & $42.1\pm 1.8$ & $ 40\,{}^{+16}_{-14}$ \\
  $\mathrm{Br}(B^0 \to K^{*0}\gamma)$ & $43.3\pm 1.5$ & $ 40\,{}^{+16}_{-14}$ \\
  $A_{CP}(B^+ \to K^{*+}\gamma)$   & $  18\pm  29$ & $-0.13\,{}^{+0.19}_{-0.19}$ \\
  $A_{CP}(B^0 \to K^{*0}\gamma)$   & $-0.1\pm 1.5$~\cite{ICHEP:raddecays} & $0.32\,{}^{+0.28}_{-0.27}$ \\
  $\Delta(K^*\gamma)$              & $ 5.2\pm 2.6$ & $ 4.1\,{}^{+2.4}_{-2.0}$ \\
\hline
  $\mathrm{Br}(B^+ \to \rho^{+}\gamma)$  & $0.98\,{}^{+0.25}_{-0.24}$ & $ 1.54\,{}^{+0.53}_{-0.48}$ \\
  $\mathrm{Br}(B^0 \to \rho^{0}\gamma)$  & $0.86\,{}^{+0.15}_{-0.14}$ & $ 0.78\,{}^{+0.26}_{-0.24}$ \\
  $A_{CP}(B^+ \to \rho^{+}\gamma)$  & $-11 \pm 33$ & $-9.4\,{}^{+2.9}_{-4.1}$ \\
  $A_{CP}(B^0 \to \rho^{0}\gamma)$  & $-44 \pm 51$ & $-9.0\,{}^{+2.7}_{-3.4}$ \\
  $\Delta(\rho\gamma)$              & $-46\,{}^{+17}_{-16}$ & $-7.9\,{}^{+5.2}_{-5.5}$ \\
\hline
  $\mathrm{Br}(B^0 \to \omega\gamma)$  & $0.44\,{}^{+0.18}_{-0.16}$ & $ 0.57\,{}^{+0.30}_{-0.25}$ \\
  $A_{CP}(B^0 \to \omega\gamma)$  &    & $-8.6\,{}^{+3.0}_{-3.6}$ \\
\hline
  $\mathrm{Br}(B_s \to \overline{K}^{*0}\gamma)$  &  & $ 1.57\,{}^{+0.60}_{-0.53}$ \\
  $A_{CP}(B_s \to \overline{K}^{*0}\gamma)$       &  & $-8.4\,{}^{+2.9}_{-3.4}$ \\
\hline
  $\mathrm{Br}(B_s \to \phi\gamma)$ & $33\pm 3$~\cite{:2012qi,ICHEP:raddecays} & $ 51\,{}^{+16}_{-14}$ \\
  $A_{CP}(B_s \to \phi\gamma)$ &  & $0$ \\
\hline
  $\mathrm{Br}(B\to X_s \gamma)$     & $341\pm 22$~\cite{Lees:2012ym} & $308\,{}^{+23}_{-26}$ \\
  $A_{CP}(B\to X_s \gamma)$          & $-1.2\pm 2.8$                  & $2.2\,{}^{+0.6}_{-2.8}$ \\
\hline
  $\mathrm{Br}(B\to X_d \gamma)$     & $14.1\pm 4.9$~\cite{delAmoSanchez:2010ae,Wang:2011sn} & $14.2\,{}^{+1.8}_{-2.9}$ \\
  $A_{CP}(B\to X_d \gamma)$          &                                                       & $-48\,{}^{+62}_{-14}$ \\
\hline \hline
\end{tabular}
\footnotesize \parbox[thb]{15.5cm} {\vspace{0.15cm} $^\dagger$ The values given here correspond to a photon-energy cut at $E_{\gamma}=1.6~{\rm GeV}$~\cite{Amhis:2012bh}.}
\end{center}
\end{table}

Within the SM, our predictions for the CP-averaged branching ratios as well as the CP and isospin asymmetries are collected in Table~\ref{tab:SMPredictions}. The theoretical uncertainties are obtained by varying the input parameters listed in the appendix within their respective ranges and adding them in quadrature, while the experimental data, if not stated otherwise, is taken from the Heavy Flavor Averaging Group~\cite{Amhis:2012bh}. For completeness, we also present in Table~\ref{tab:SMPredictions} our predictions for the inclusive $B\to X_s \gamma$ and $B\to X_d \gamma$ decays, the theoretical framework of which can be found in Refs.~\cite{Misiak:2006ab,Misiak:2006zs,Benzke:2010js,Crivellin:2011ba} for the branching ratios, and in Refs.~\cite{Kagan:1998bh,Hurth:2003dk,Benzke:2010tq} for the direct CP asymmetries.

The main theoretical uncertainties are stemming from the transition form factor $T_1(0)$ for the branching ratios, the variation of renormalization scales $\mu_b$, $\mu_{hc}$ for the asymmetries in $B\to \rho\gamma$ modes, and the first Gegenbauer moment $a_1^{\perp}$ for the asymmetries in $B\to K^{\ast}\gamma$ modes. We have added a global $15\%$ uncertainty in all exclusive observables to account for non-factorizable effects, not yet included in the QCDF framework. We note that, taking into account their respective uncertainties, the predictions for all observables are in good or very good agreement with the data, the only tension appearing in the isospin asymmetry in $B\to\rho\gamma$, which has however a rather large experimental uncertainty.

The agreement for the inclusive $B\to X_{s}\gamma$ (and $B\to X_{d}\gamma$) modes implies very stringent constraints on various NP models~\cite{B2Vgreview,B2Vg-NP,B2Vg-NP-2,Crivellin:2011ba}. For the direct CP asymmetries, on the other hand, due to the hadronic component of the photon, there are quite large theoretical uncertainties, lowering the predictive power of these observables~\cite{Benzke:2010tq}.

In the numerical analysis, we impose the experimental constraints in the following way: each point in the A2HDM parameter space corresponds to a theoretical range, constructed as the prediction for this observable in that point together with the corresponding theory error. If this range has overlap with the $2\sigma$ range of the measurement, we consider the point allowed. In that procedure, the relative theory uncertainty is assumed constant over the parameter space. This is a reasonable assumption, since the main theoretical uncertainties are due to the hadronic input parameters, common to both the SM and the NP contributions. In order to obtain constraints on the charged-scalar Yukawa couplings $|\varsigma_{u,d}|$, we vary the remaining parameters in the ranges $m_{H^{\pm}}\in[80,500]~{\rm GeV}$ and $\phi\in[0,360^\circ]$.

\subsection{$B\to X_{s,d}\,\gamma$ decays within the A2HDM}
\label{Sec:B2XsdG}

Since the inclusive $B\to X_{s,d}\,\gamma$ decay amplitudes are proportional to the effective Wilson coefficient $C_{7}^{\mathrm{eff}}(\mu_b)$, to facilitate the following discussions, we can decompose it in such a way that the dependence on the couplings $\varsigma_u$ and $\varsigma_d$ becomes manifest. In the LO approximation and at the scale $\mu_b=2.5~\mathrm{GeV}$, we have numerically
\begin{equation}\label{Eq:C7mub}
 C_7^{\mathrm{eff}}(\mu_b)\; =\; C_{7,\mathrm{SM}}^{\mathrm{eff}}(\mu_b)\,\biggl\{1 - 0.29\:\varsigma_u^*\varsigma_d\,\left(\frac{200~\mathrm{GeV}}{m_{H^{\pm}}}\right)^2 + 0.05\: |\varsigma_u|^2\,\left(\frac{200~\mathrm{GeV}}{m_{H^{\pm}}}\right)^2 \biggr\}\,,
\end{equation}
from which we can see that, even for comparable $|\varsigma_u|$ and $|\varsigma_d|$, a stringent constraint on the combination $\varsigma_u^*\varsigma_d$ is expected from these decays.

\begin{figure}[thb]
\centering
\includegraphics[width=15cm]{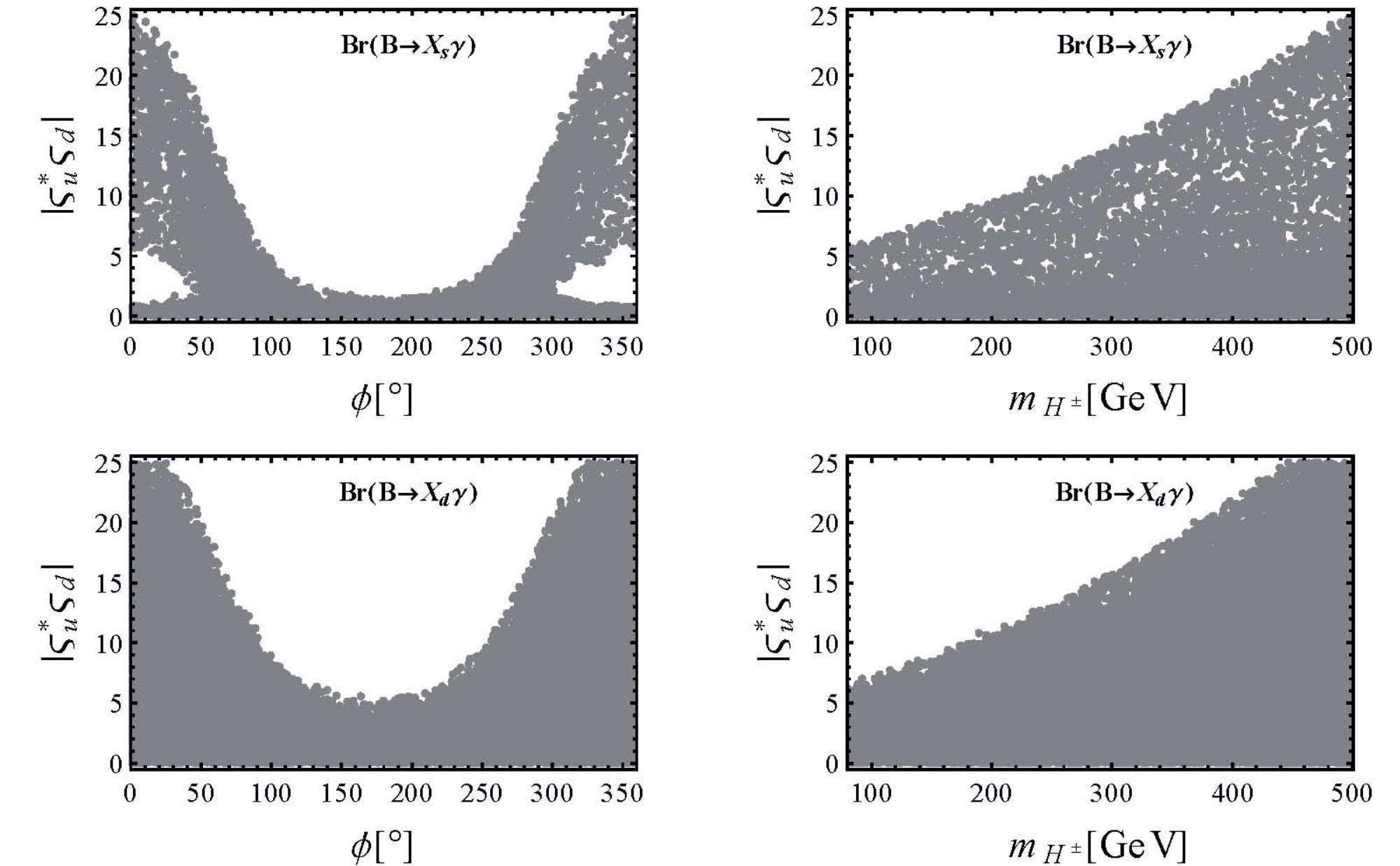}
\caption{\label{fig:B2XsdG-Complex} \small Constraints on the charged-scalar couplings $\varsigma_u$ and $\varsigma_d$ from inclusive $B \to X_{s,d}\,\gamma$ decays, plotted in the planes $|\varsigma_u^*\varsigma_d|-\phi$~(left) and $|\varsigma_u^*\varsigma_d|-m_{H^{\pm}}$~(right). The charged-scalar mass $m_{H^{\pm}}$ and the relative phase $\phi$ are varied in the ranges $[80,500]~{\rm GeV}$ and $[0,360^\circ]$, respectively.}
\end{figure}

\begin{figure}[thb]
\centering
\includegraphics[width=15cm]{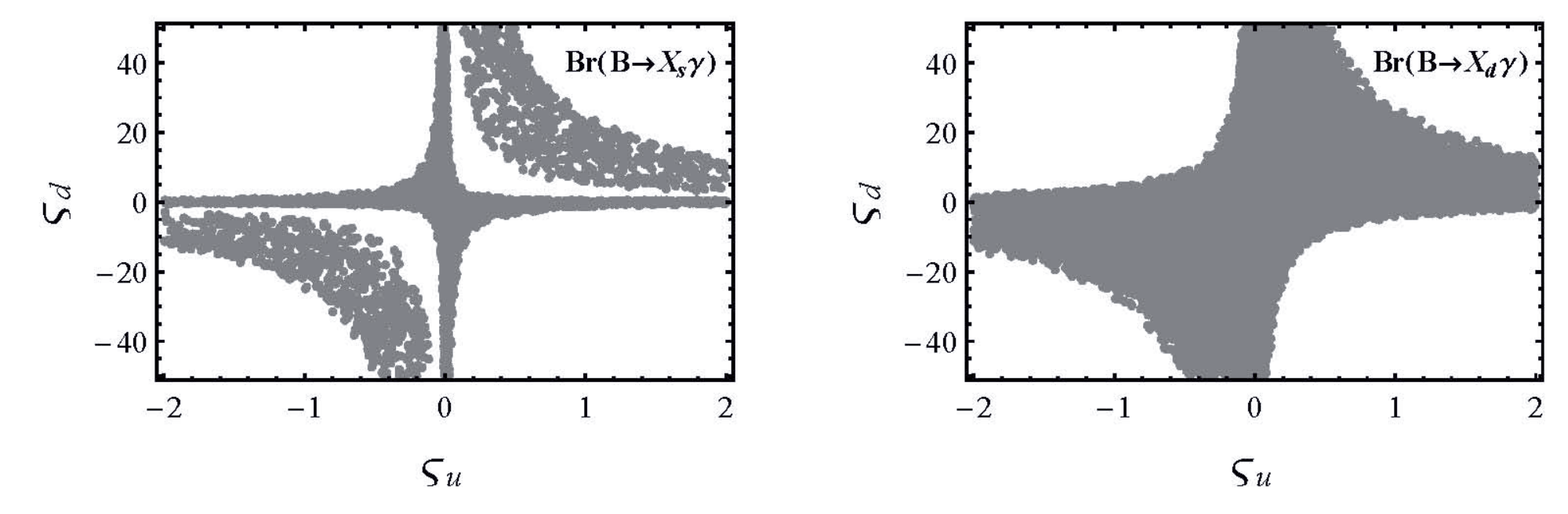}
\caption{\label{fig:B2XsdG-Real} \small Constraints as in Fig.~\ref{fig:B2XsdG-Complex}, but for real couplings plotted in the $\varsigma_u-\varsigma_d$ plane.}
\end{figure}

Following the discussions in Refs.~\cite{Jung:2010ik,Jung:2010ab}, we show in Figs.~\ref{fig:B2XsdG-Complex} and \ref{fig:B2XsdG-Real} the updated constraints on the charged-scalar couplings $\varsigma_u$ and $\varsigma_d$, corresponding to the complex and the real case, respectively. We make the following observations:
\begin{enumerate}

\item[$\bullet$] The current experimental data on $\mathrm{Br}(B\to X_s\gamma)$ gives the most stringent constraint on the couplings $\varsigma_u$ and $\varsigma_d$, for both the complex and real cases. For $\phi\sim\pi$ only a small allowed region remains due to the constructive interference between the SM and the NP contributions, similar to the type-II 2HDM, while for $\phi\sim 0$ the interference between them is destructive, and there are two allowed regions, corresponding to relatively small NP influence~(the lower region) and the case where it is about twice the size of the SM contribution~(the upper region).

\item[$\bullet$] This implies for real couplings that while simultaneously large values for real $\varsigma_u$ and $\varsigma_d$ with the same signs remain allowed, they are excluded for different signs.

\item[$\bullet$] Taking into account the constraint from the branching ratio, the CP asymmetry does not constrain the NP parameter space further.

\item[$\bullet$] As is obvious from Eq.~(\ref{Eq:C7mub}), the combination $|\varsigma_u^*\varsigma_d|$ is strongly correlated with the charged-scalar mass, and large values are only allowed for large $m_{H^{\pm}}$.

\end{enumerate}

\begin{figure}[thb]
\centering
\includegraphics[width=15cm]{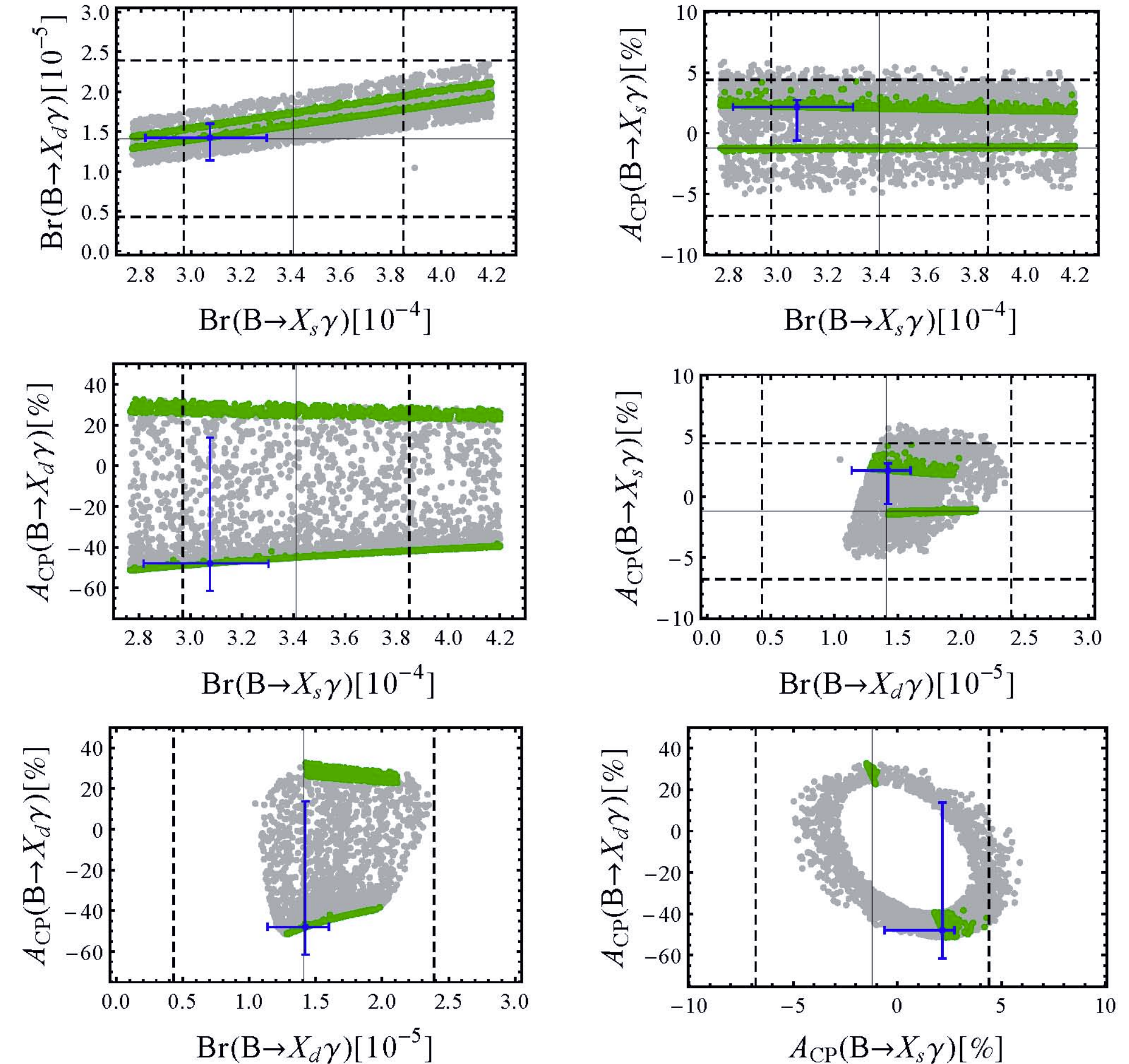}
\caption{\label{fig:B2XsG-B2XdG-correlation} \small Correlation plots between the observables in inclusive $B\to X_{s,d}\,\gamma$ decays. The green~(dark grey) and grey~(light grey) regions correspond to the real and complex couplings, respectively. The solid and dashed lines denote the experimental central value and the corresponding $2\sigma$ range, while the SM predictions with $1\sigma$ error bars are shown by the blue~(dark) cross.}
\end{figure}

Since the branching ratio $\mathrm{Br}(B\to X_s \gamma)$ is a key observable, it is interesting to investigate its correlations with the other observables. Furthermore, as its effect on the A2HDM is not as simple as for the type-II 2HDM, but is merely to strongly correlate the different parameters, correlations between observables are significantly affected when imposing the corresponding constraint. We show both in Fig.~\ref{fig:B2XsG-B2XdG-correlation}, constructed as described above. As the uncertainties of the other observables are mostly independent of the one from $B\to X_s\gamma$, the cross for the hadronic uncertainties is to be applied to each of these points. Apart from the trivial one between $\mathrm{Br}(B\to X_{s,d}\gamma)$, we observe mild correlations for most observables, the exception being the two direct CP asymmetries. As concluded already in \cite{Jung:2010ab}, here an improvement in the experimental precision might lead to interesting insights, it would however have to be complemented by theoretical progress, given the theoretical error for these observables.

\subsection{$B\to K^*\gamma$ decays within the A2HDM}
\label{Sec:B2KVG}

For the exclusive $B\to V\gamma$ decays, the decay amplitudes are also proportional to the coefficient $C_{7}^{\mathrm{eff}}(\mu_b)$ through the quantity ${\cal C}_7^{(i)}$ defined in Eq.~(\ref{CalC7}). Accordingly, the general observations for the inclusive $B\to X_{s,d}\,\gamma$ decays apply here as well. However, as we have the isospin asymmetry as an additional observable, different constraints on the model parameters are expected. In this subsection, we shall discuss $B\to K^*\gamma$ decays.

\begin{figure}[thb]
\centering
\includegraphics[width=15cm]{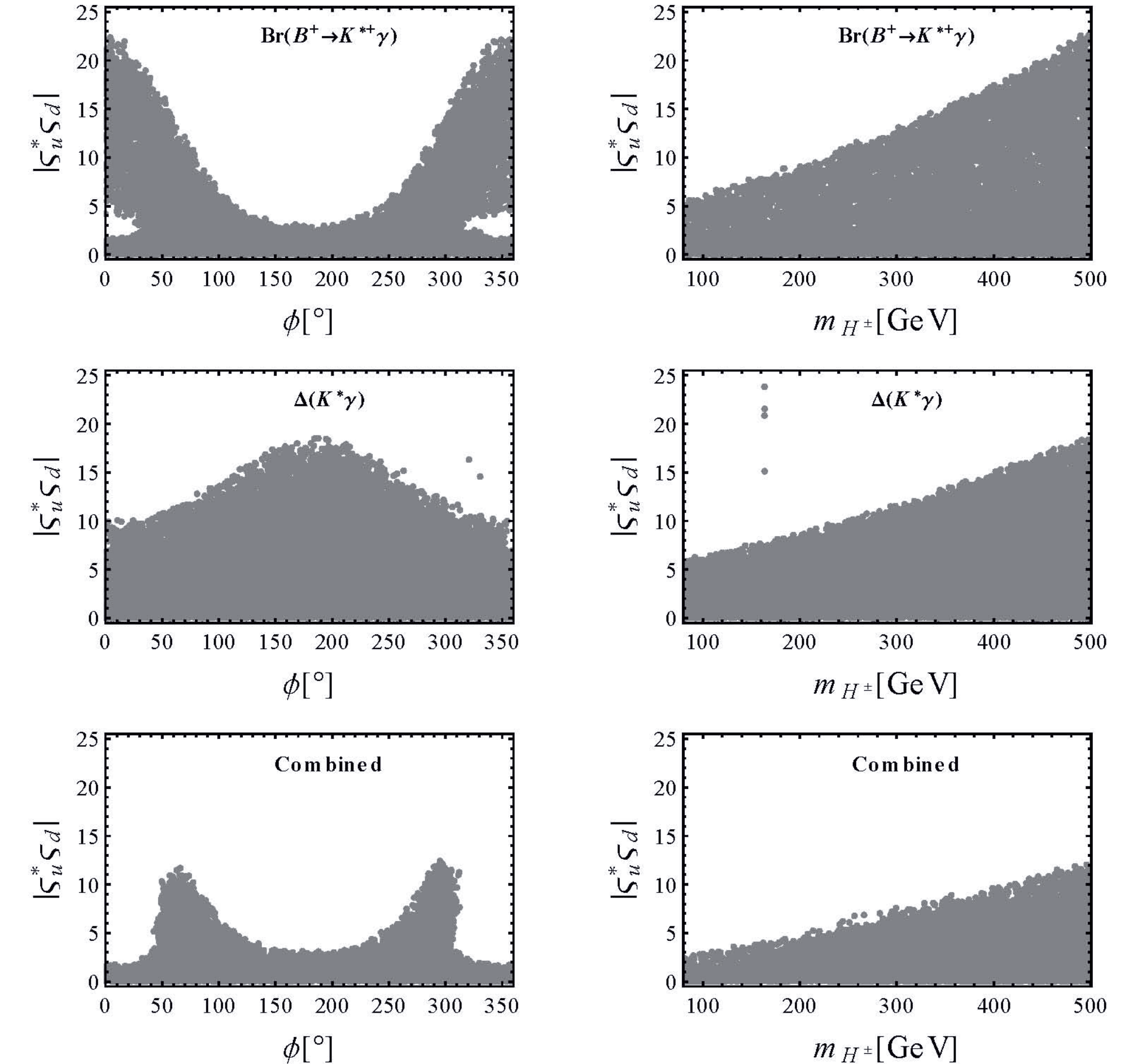}
\caption{\label{fig:B2KVG-Complex} \small Constraints from $B \to K^* \gamma$ decays, plotted in the planes $|\varsigma_u^*\varsigma_d|-\phi$~(left) and $|\varsigma_u^*\varsigma_d|-m_{H^{\pm}}$~(right). The other captions are the same as in Fig.~\ref{fig:B2XsdG-Complex}.}
\end{figure}

\begin{figure}[thb]
\centering
\includegraphics[width=15cm]{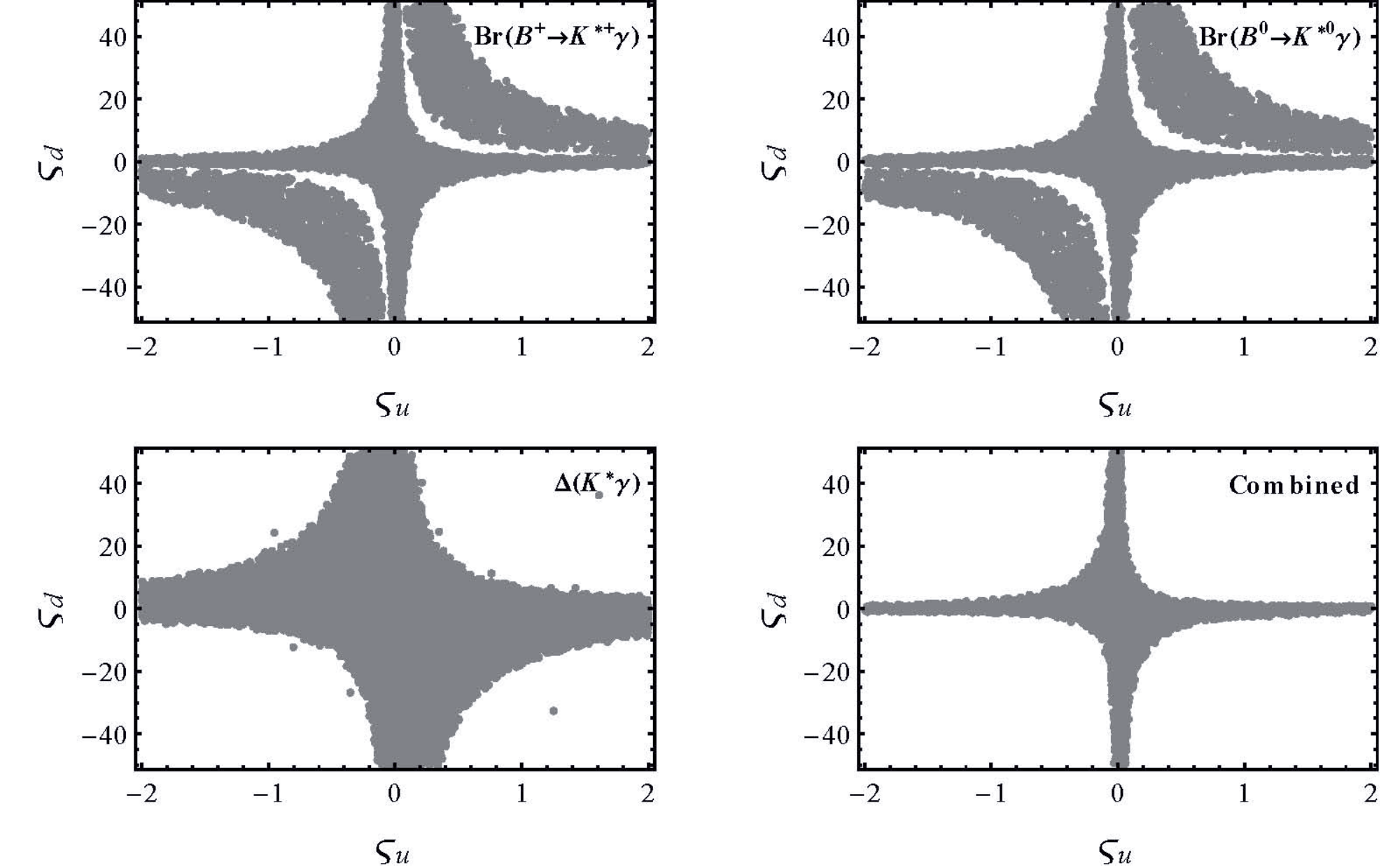}
\caption{\label{fig:B2KVG-Real} \small Constraints as in Fig.~\ref{fig:B2KVG-Complex}, but for real couplings plotted in the $\varsigma_u-\varsigma_d$ plane.}
\end{figure}

Imposing the current experimental data on $B\to K^*\gamma$ decays as constraints, we show in Figs.~\ref{fig:B2KVG-Complex} and \ref{fig:B2KVG-Real} the allowed regions for the charged-scalar couplings, corresponding to the complex and the real case, respectively. From these plots, the following observations are made:
\begin{enumerate}

\item[$\bullet$] Constraints on the model parameters from the two CP-averaged branching ratios $\mathrm{Br}(B^+\to K^{*+}\gamma)$ and $\mathrm{Br}(B^0\to K^{*0}\gamma)$ are quite similar to those from $\mathrm{Br}(B\to X_s\gamma)$, since their decay amplitudes are all dominated by the coefficient $C_{7}^{\mathrm{eff}}(\mu_b)$.

\item[$\bullet$] Since there are still large theoretical and experimental uncertainties for the direct CP asymmetries in these decays, again almost no constraints can be obtained from these observables, which is the reason why they are not shown here.

\item[$\bullet$] The isospin asymmetry $\Delta(K^*\gamma)$ varies like $1/C_{7}^{\mathrm{eff}}(\mu_b)$ to first order, and consequently has a different dependence on the relative phase $\phi$, as shown in the third plot of Fig.~\ref{fig:B2KVG-Complex}. This allows to exclude the large same-sign solutions allowed by the branching ratios.

\item[$\bullet$] Once constraints from the branching ratios and the isospin asymmetry are combined, the allowed parameter space is therefore severely reduced, as shown in the last two plots of Fig.~\ref{fig:B2KVG-Complex} for complex, and the last plot of Fig.~\ref{fig:B2KVG-Real} for real couplings.

\end{enumerate}

\begin{figure}[thb]
\centering
\includegraphics[width=15cm]{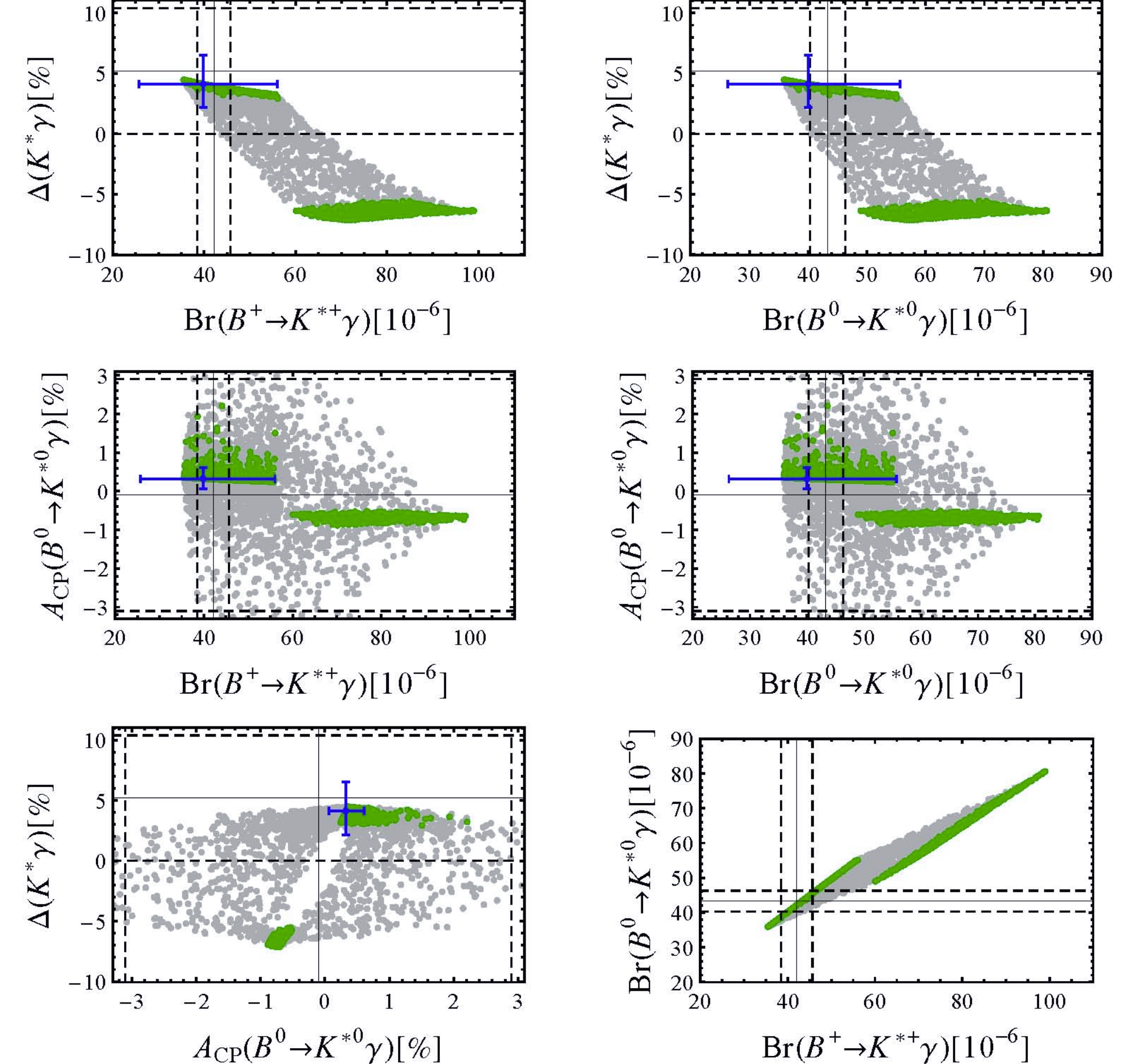}
\caption{\label{fig:B2XsG-B2KVG-correlation-2} \small Correlation plots between the observables in $B\to K^*\gamma$ decays, see also Fig.~\ref{fig:B2XsG-B2XdG-correlation}.}
\end{figure}

In Fig.~\ref{fig:B2XsG-B2KVG-correlation-2} the correlations between different observables in $B\to K^*\gamma$ decays are shown. The strongest non-trivial ones are between the isospin asymmetry and the branching ratios; these are however not very effective within the experimentally allowed range. Very large values for $\Delta(K^*\gamma)$ correspond to the case where a strong cancellation between the SM and the NP contributions to $C_{7}^{\mathrm{eff}}$ occurs, making the remaining parts, such as the annihilation and spectator-scattering contributions, relatively important.

Thus, it is concluded that the data from $B\to K^*\gamma$ decays, especially the isospin asymmetry, constrain the parameter space in a way complementary to $B\to X_s\gamma$. A further improved measurement of this quantity will therefore be important in constraining NP.

\subsection{$B\to \rho\gamma$ decays within the A2HDM}
\label{Sec:B2RhoG}

For the exclusive $B\to \rho \gamma$ decays, since the CKM factors $\lambda_u^{(d)}$ and $\lambda_t^{(d)}$ are comparable in magnitude, both of the two decay amplitudes ${\cal C}_7^{(u)}$ and ${\cal C}_7^{(t)}$ contribute effectively. This feature makes these decays particularly interesting in constraining the CKM unitarity triangle and searching for physics beyond the SM~\cite{Muheim:2008vu,B2Vg-LCSR,Bosch:2001gv,B2Vg-NP,B2Vg-NP-2}.

\begin{figure}[thb]
\centering
\includegraphics[width=15cm]{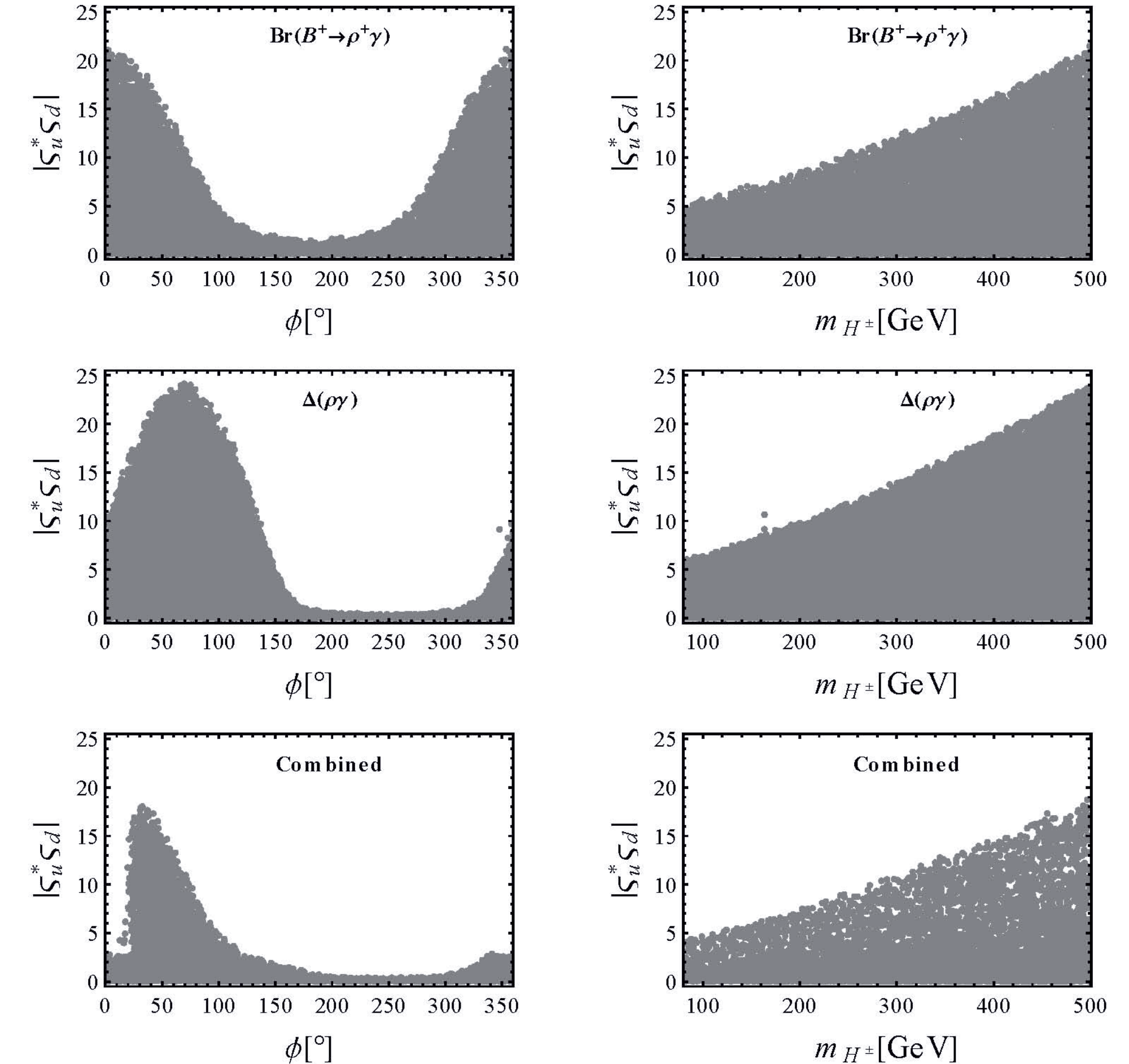}
\caption{\label{fig:B2RhoG-Complex} \small  Constraints from $B \to \rho\gamma$ decays, plotted in the planes $|\varsigma_u^*\varsigma_d|-\phi$~(left) and $|\varsigma_u^*\varsigma_d|-m_{H^{\pm}}$~(right). The other captions are the same as in Fig.~\ref{fig:B2XsdG-Complex}.}
\end{figure}

\begin{figure}[thb]
\centering
\includegraphics[width=15cm]{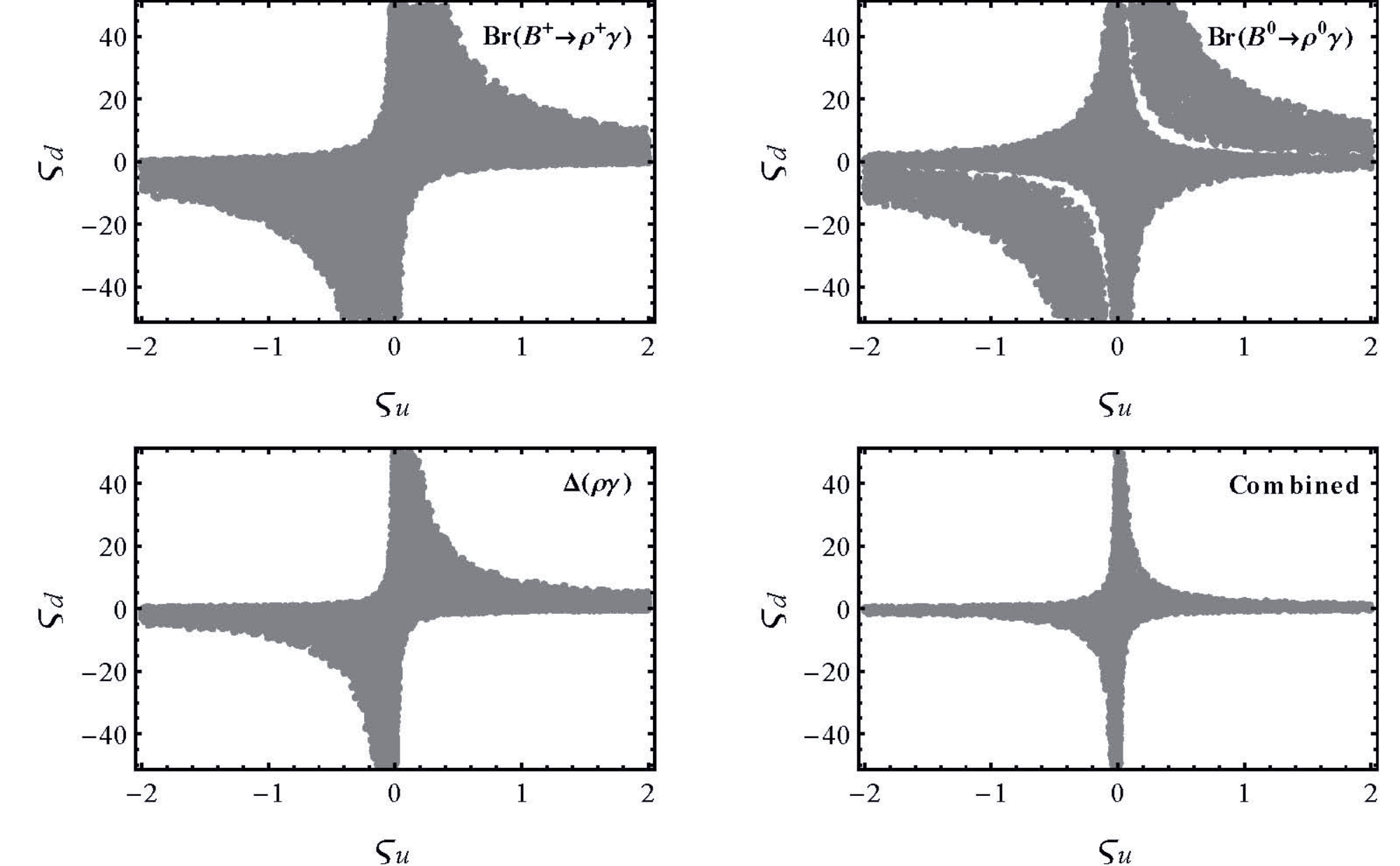}
\caption{\label{fig:B2RhoG-Real} \small Constraints as in Fig.~\ref{fig:B2RhoG-Complex}, but for real couplings plotted in the $\varsigma_u-\varsigma_d$ plane.}
\end{figure}

\begin{figure}[thbp]
\centering
\includegraphics[width=15cm]{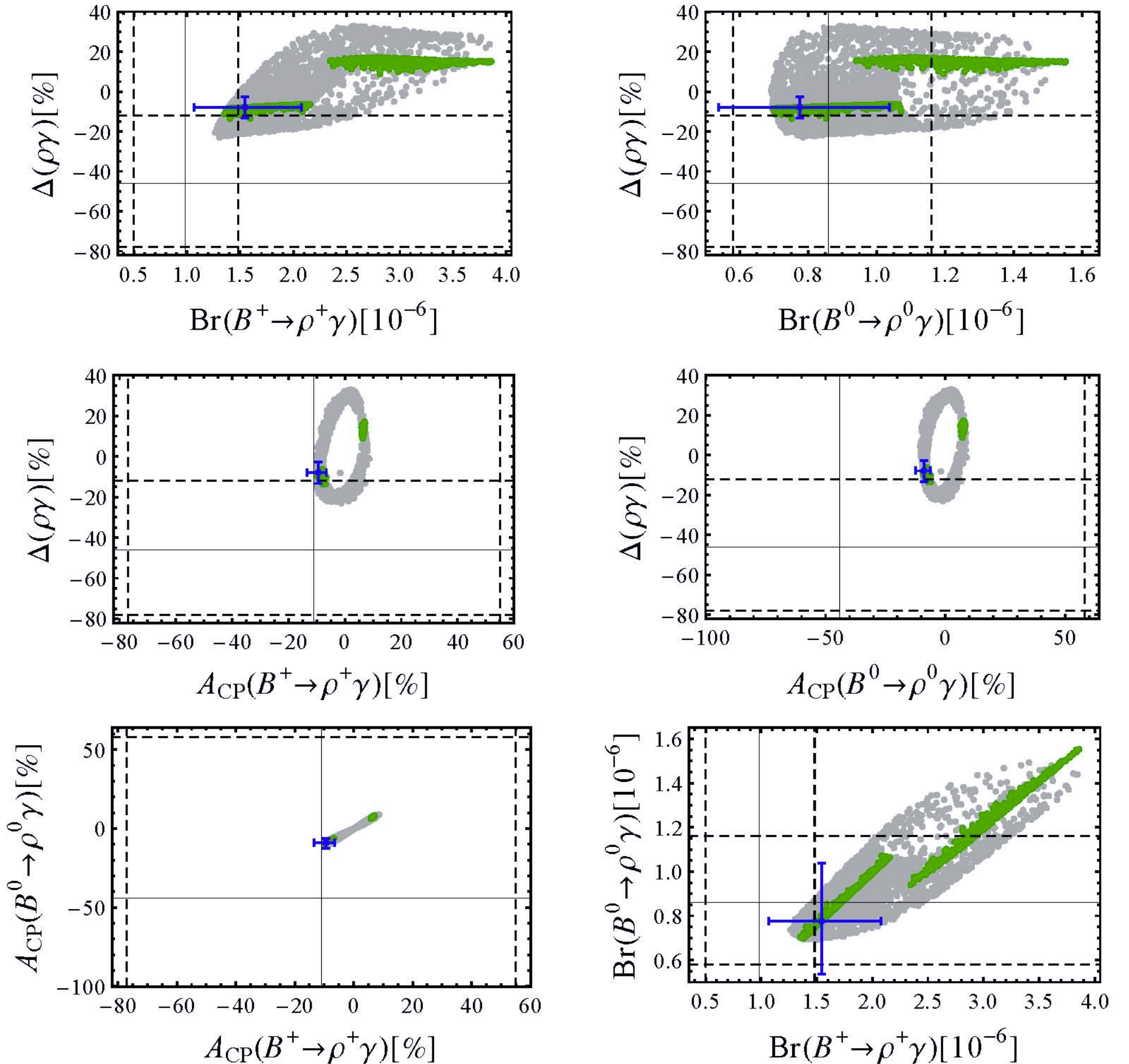}
\caption{\label{fig:B2XsG-B2RhoG-correlation-2} \small Correlation plots between the observables in $B\to\rho\gamma$ decays, see also Fig.~\ref{fig:B2XsG-B2XdG-correlation}.}
\end{figure}

\begin{figure}[thbp]
\centering
\includegraphics[width=7.2cm]{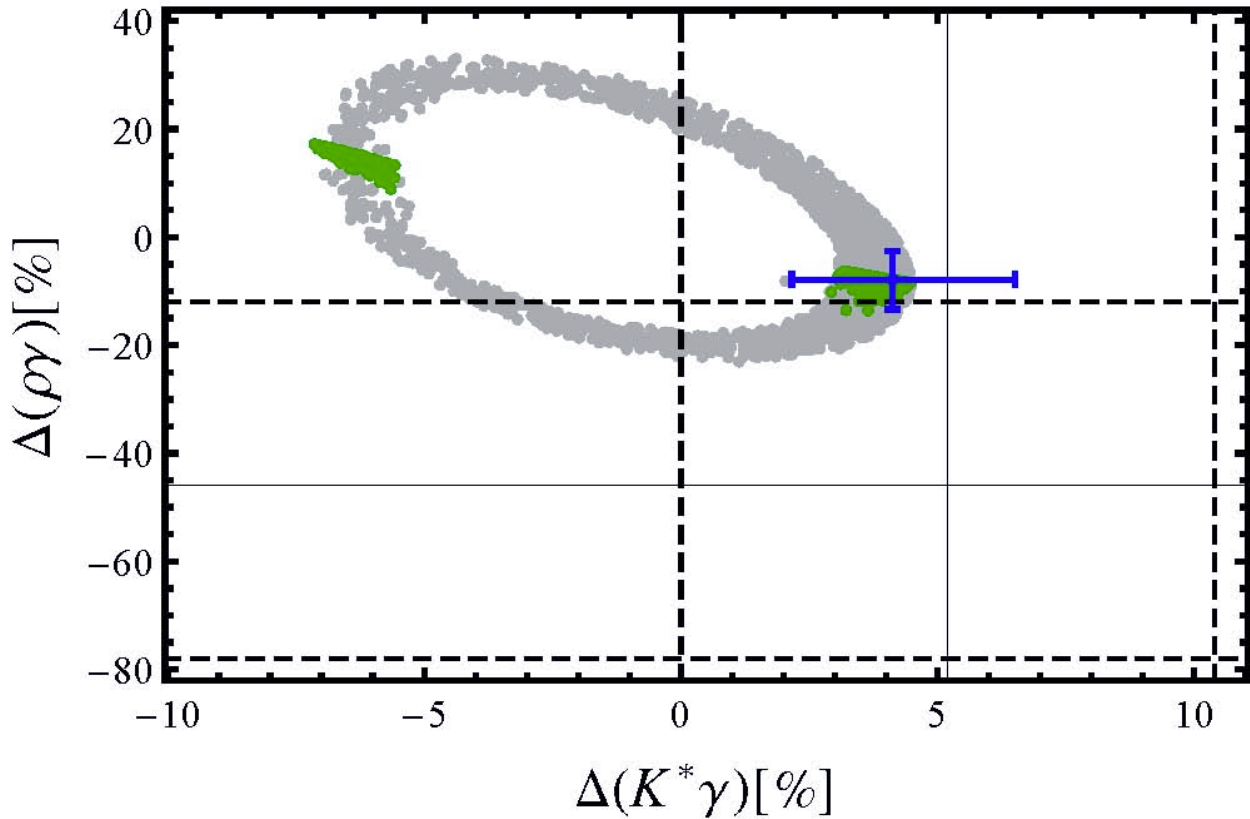}
\caption{\label{fig:DeltaKVG_DeltaRhoG} \small Correlation plot between $\Delta(K^*\gamma)$ and $\Delta(\rho\gamma)$, see also Fig.~\ref{fig:B2XsG-B2XdG-correlation}.}
\end{figure}

Focusing on the A2HDM effect, we show in Figs.~\ref{fig:B2RhoG-Complex} and \ref{fig:B2RhoG-Real} the current constraints on the couplings $\varsigma_u$ and $\varsigma_d$ from these decays, corresponding to the complex and the real case, respectively. Similarly to the discussions for $B\to K^*\gamma$ decays, the correlations between the observables in $B\to \rho\gamma$ decays are shown in Fig.~\ref{fig:B2XsG-B2RhoG-correlation-2}. Furthermore, we show in Fig.~\ref{fig:DeltaKVG_DeltaRhoG} the correlation between the isospin asymmetries in $B\to K^*\gamma$ and $B\to \rho\gamma$. From these plots, we make the following observations: \begin{enumerate}

\item[$\bullet$] From the two branching ratios, currently the charged mode $\mathrm{Br}(B^+\to \rho^+ \gamma)$ gives a much stronger constraint. The direct CP asymmetries are again found not to be able to put any constraints on the model parameters.

\item[$\bullet$] The isospin asymmetry $\Delta(\rho\gamma)$ again shows a different dependence on the relative phase $\phi$ than the branching ratios, but also than $\Delta(K^*\gamma)$. This is due to the contribution from the extra term proportional to $\lambda_u^{(d)}$, which is associated with a different weak phase $\arg(V_{ub})$.

\item[$\bullet$] The combined constraint from the branching ratios and the isospin asymmetry reduces significantly the parameter space allowed by the individual observables, showing the important role played by $\Delta(\rho\gamma)$.

\item[$\bullet$] Even with the constraint from $\mathrm{Br}(B\to X_s\gamma)$ imposed, very large values for the direct CP and isospin asymmetries remain allowed, which is again due to the relatively large term $\sim\lambda_u^{(d)}$. This part also includes the formally power-suppressed weak annihilation contribution, which is in this case enhanced by the large Wilson coefficients $C_{1,2}$.

\item[$\bullet$] With the constraint from $\mathrm{Br}(B\to X_s\gamma)$ taken into account, even in the A2HDM the very large central value of $\Delta(\rho\gamma)$ cannot be reproduced. Its enhancement would imply smaller CP asymmetries in both decay modes.

\item[$\bullet$] There is a strong correlation between the two isospin asymmetries, and relatively large deviations from the SM values remain allowed by the data in the A2HDM, although not as large as the present central value of $\Delta(\rho\gamma)$.

\end{enumerate}

Thus, similarly to $B\to K^*\gamma$, the isospin asymmetry $\Delta(\rho\gamma)$ is a very important observable in constraining the charged-scalar Yukawa couplings, and a more precise measurement will yield even stronger constraints. A confirmation of the present central value to higher precision would challenge the SM as well as most 2HDMs, including the A2HDM.

\subsection{Other $B\to V\gamma$ decays within the A2HDM}
\label{Sec:OtherModes}

In this subsection, we discuss the $b\to d$ decay modes $B_d\to \omega \gamma$ and $B_s\to \overline{K}^{*0}\gamma$, as well as the $b\to s$ one $B_s \to \phi \gamma$. The branching ratio of the latter has very recently been measured precisely~\cite{:2012qi,ICHEP:raddecays}, while for $B_d\to \omega\gamma$ the uncertainties remain rather large, and thus only very loose constraints on the A2HDM parameters can be obtained. Since neither the CP asymmetries for these modes nor observables for $B_s\to \overline{K}^*\gamma$ have been measured so far, we predict them within the SM and in the A2HDM. The remaining $B\to V\gamma$ modes are predicted to be very tiny within the QCDF approach, $\sim \mathcal{O}(10^{-10})$~\cite{Li:2003kz}. An observed enhancement could imply either NP or a breakdown of the method, as these are pure annihilation modes. A quantitative discussion does not seem appropriate in this case, as the A2HDM does not imply large enhancements.

\begin{figure}[thb]
\centering
\includegraphics[width=15cm]{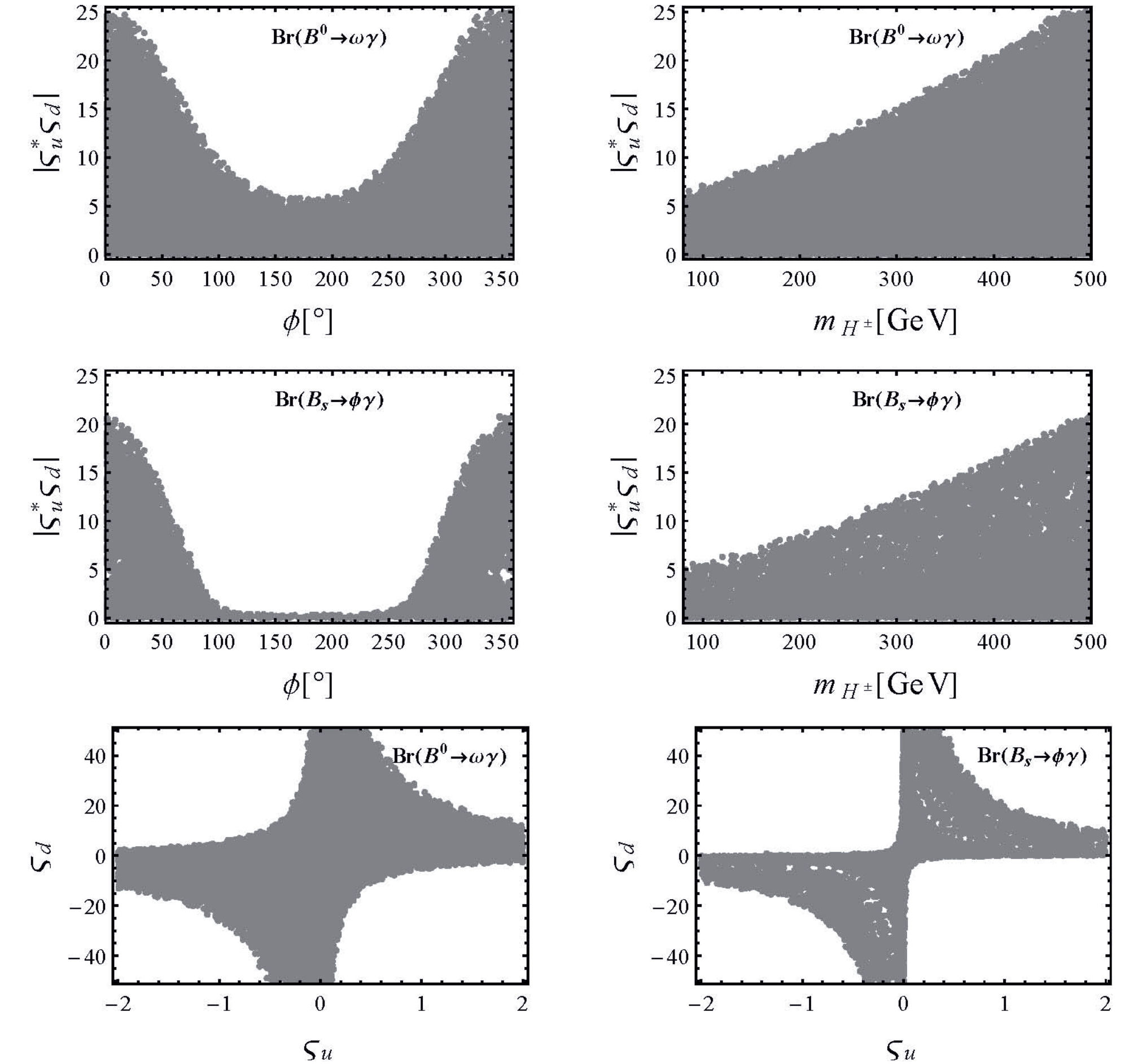}
\caption{\label{fig:OtherModes-Complex} \small Constraints from $B_d\to \omega \gamma$ and $B_s \to \phi \gamma$. The upper four and the lower two plots correspond to the complex and the real couplings, respectively.}
\end{figure}

In Fig.~\ref{fig:OtherModes-Complex} we show the constraints from $B_s\to\phi\gamma$ and $B_d \to \omega\gamma$, where the upper four and the lower two plots correspond to the complex and the real couplings, respectively. We note that $\mathrm{Br}(B_s\to\phi\gamma)$ is beginning to give competitive constraints compared to the branching ratios of the previously discussed modes.

\begin{figure}[thb]
\centering
\includegraphics[width=15cm]{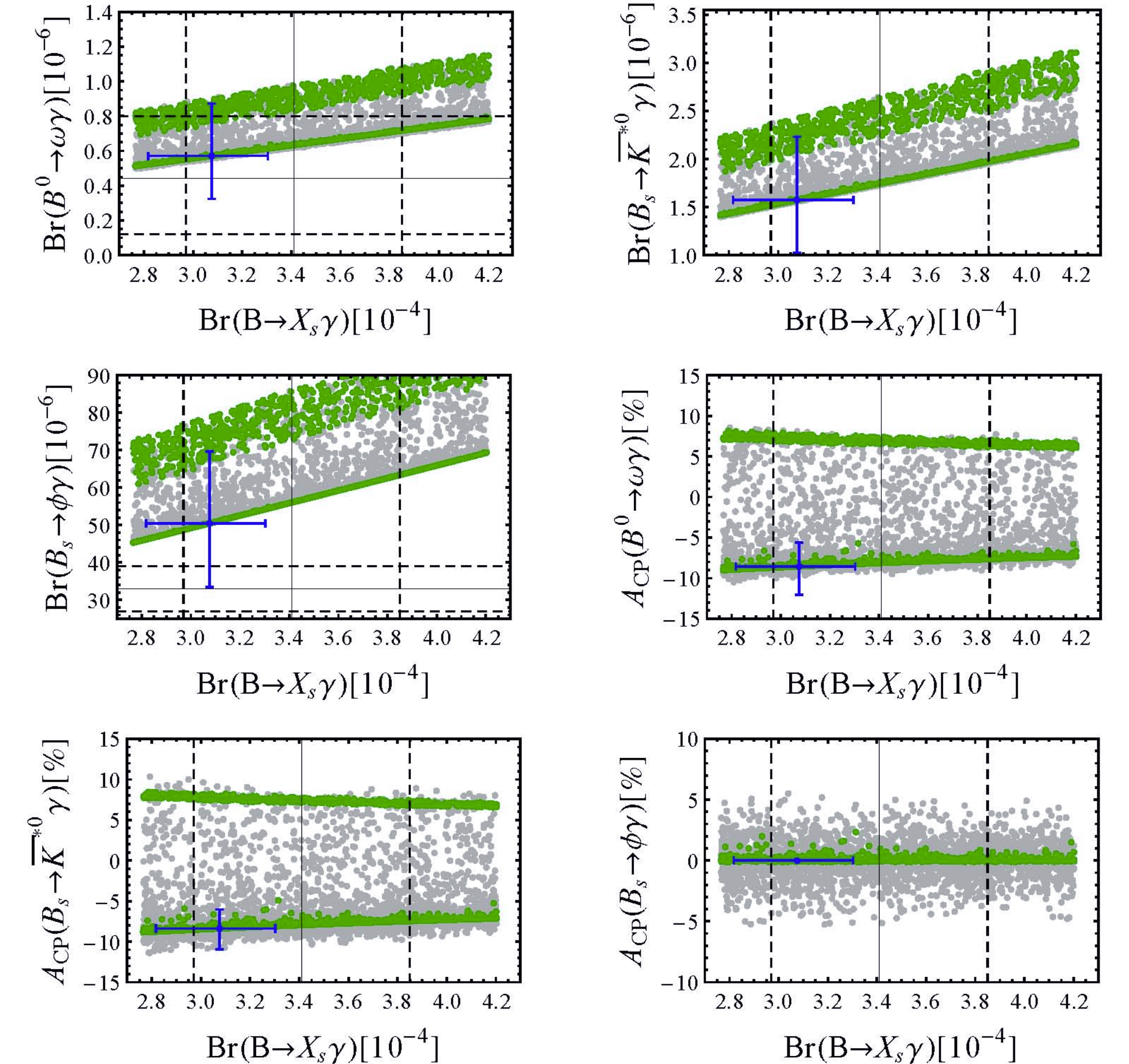}
\caption{\label{fig:B2XsG-OtherModes-correlation} \small Correlation plots between $\mathrm{Br}(B\to X_s \gamma)$ and the observables in $B_d\to \omega \gamma$, $B_s\to \overline{K}^{*0}\gamma$ and $B_s \to \phi \gamma$ decays, see also Fig.~\ref{fig:B2XsG-B2XdG-correlation}.}
\end{figure}

\begin{figure}[thb]
\centering
\includegraphics[width=15cm]{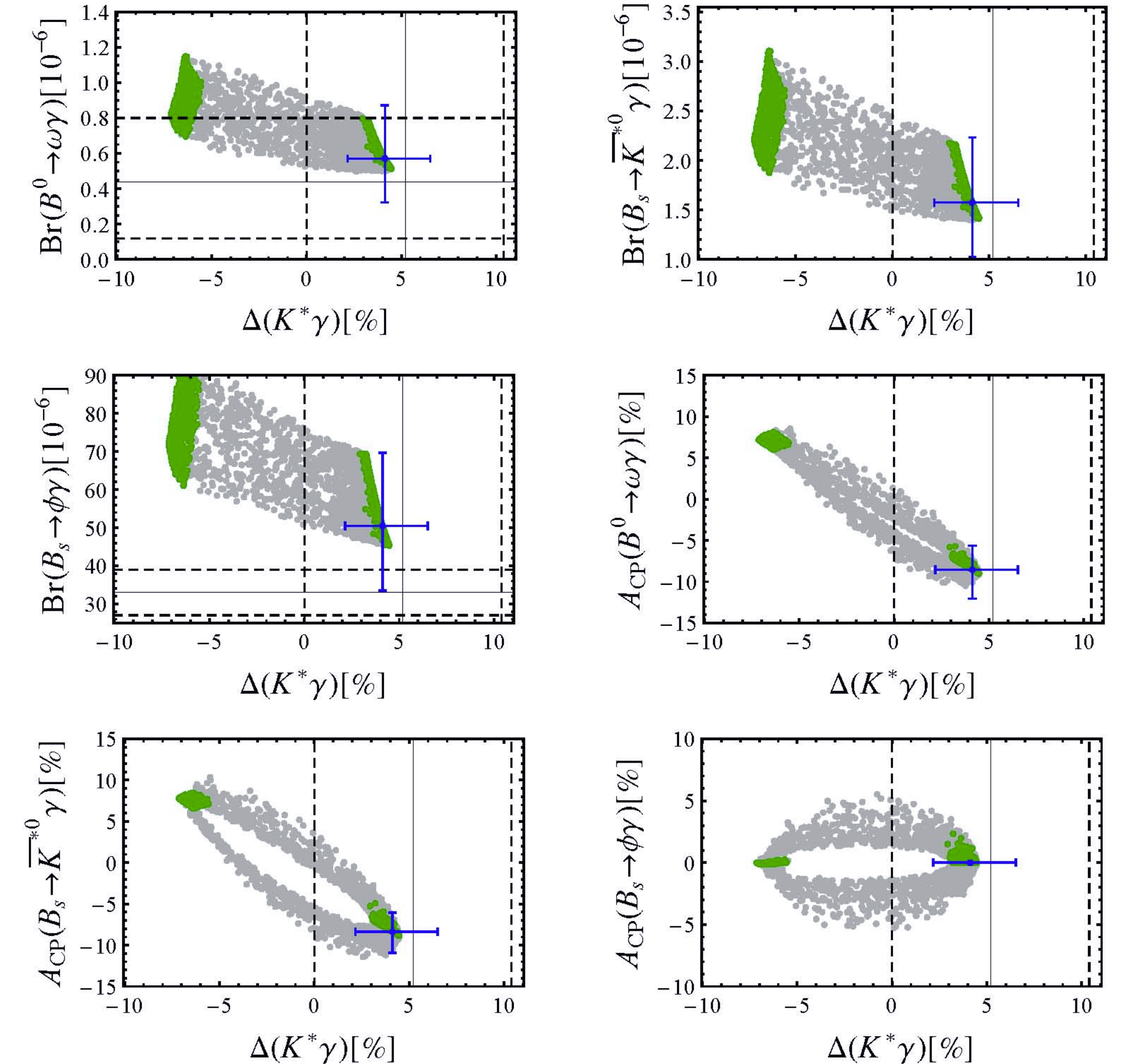}
\caption{\label{fig:DeltaKVG-OtherModes-correlation} \small Correlation plots between $\Delta(K^*\gamma)$ and the observables in $B_d\to \omega \gamma$, $B_s\to \overline{K}^{*0}\gamma$ and $B_s \to \phi \gamma$ decays, see also Fig.~\ref{fig:B2XsG-B2XdG-correlation}.}
\end{figure}

\begin{figure}[thb]
\centering
\includegraphics[width=15cm]{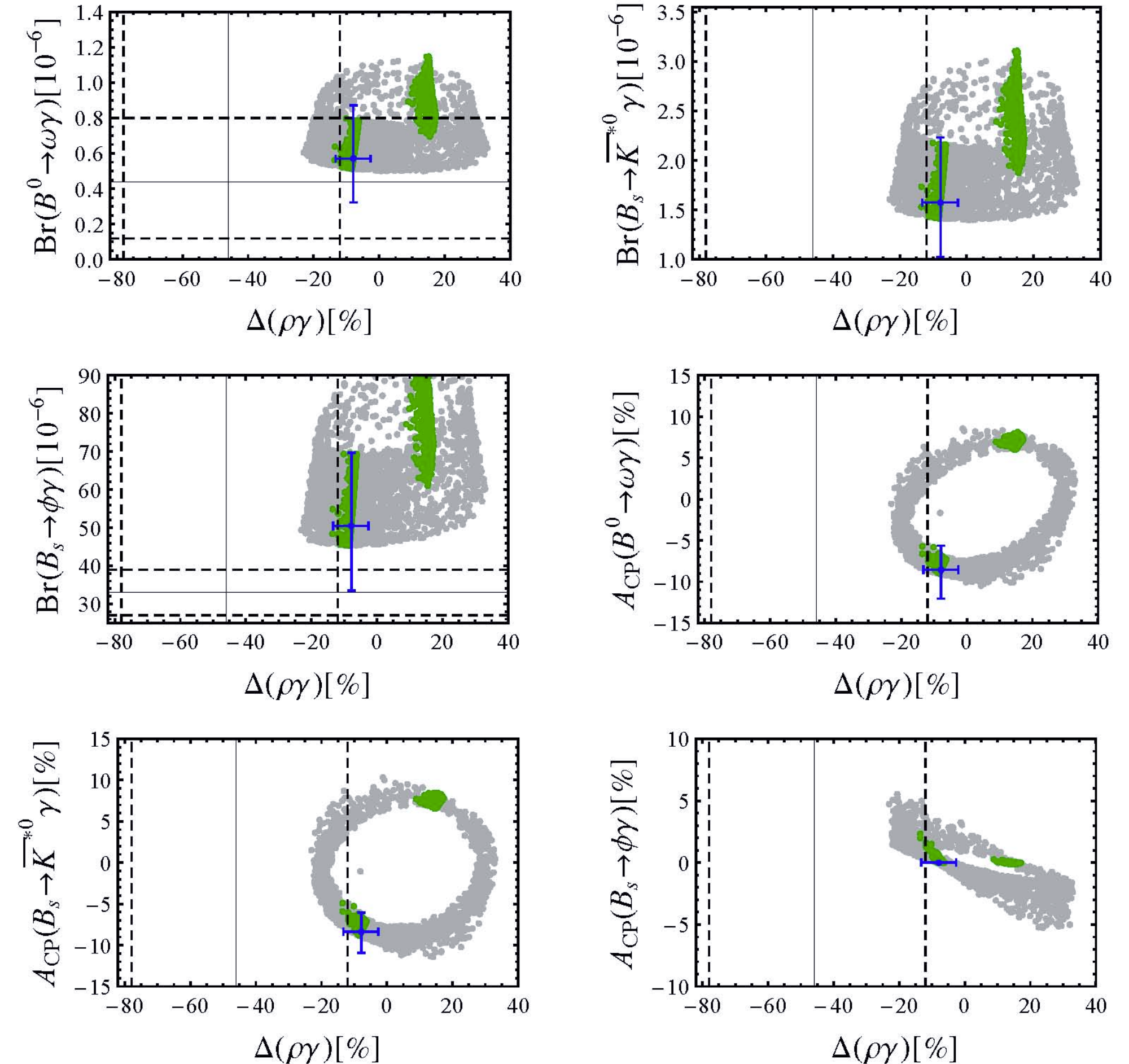}
\caption{\label{fig:DeltaRhoG-OtherModes-correlation} \small Correlation plots between $\Delta(\rho\gamma)$ and the observables in $B_d\to \omega \gamma$, $B_s\to \overline{K}^{*0}\gamma$ and $B_s \to \phi \gamma$ decays, see also Fig.~\ref{fig:B2XsG-B2XdG-correlation}.}
\end{figure}

From the previous discussions, we already know that $\mathrm{Br}(B\to X_s \gamma)$, $\Delta(K^*\gamma)$ and $\Delta(\rho\gamma)$ impose strong, complementary constraints on the A2HDM parameters. Thus, we show in Figs.~\ref{fig:B2XsG-OtherModes-correlation}-\ref{fig:DeltaRhoG-OtherModes-correlation} the correlations between these three observables and the ones in the modes presently discussed, including the ones not yet measured. These can be tested in the near future, given the expected experimental progress due to LHCb and the next-generation flavour factories.

Importantly, for all of the unmeasured CP asymmetries values very different from the SM prediction remain allowed. This observation is of special interest for the decay $B_s\to\phi\gamma$, because in this case the SM prediction of a tiny asymmetry is almost unaffected by hadronic uncertainties (see also Ref.~\cite{Muheim:2008vu}). This fact makes it a key observable not only for the A2HDM, but also for every model introducing new weak phases in $b\to s$ transitions.

\section{Conclusions}
\label{Sec:conclusion}

The A2HDM, characterized by the alignment of Yukawa matrices in flavour space, guarantees the absence of tree-level FCNCs, while introducing at the same time new sources of CP violation in the charged-scalar couplings to fermions, implying potentially large effects in many low-energy processes. In this paper, employing the QCDF approach, we have studied the exclusive radiative $B\to V\gamma$ decays within this specific NP model.

\begin{figure}[thb]
\centering
\includegraphics[width=15cm]{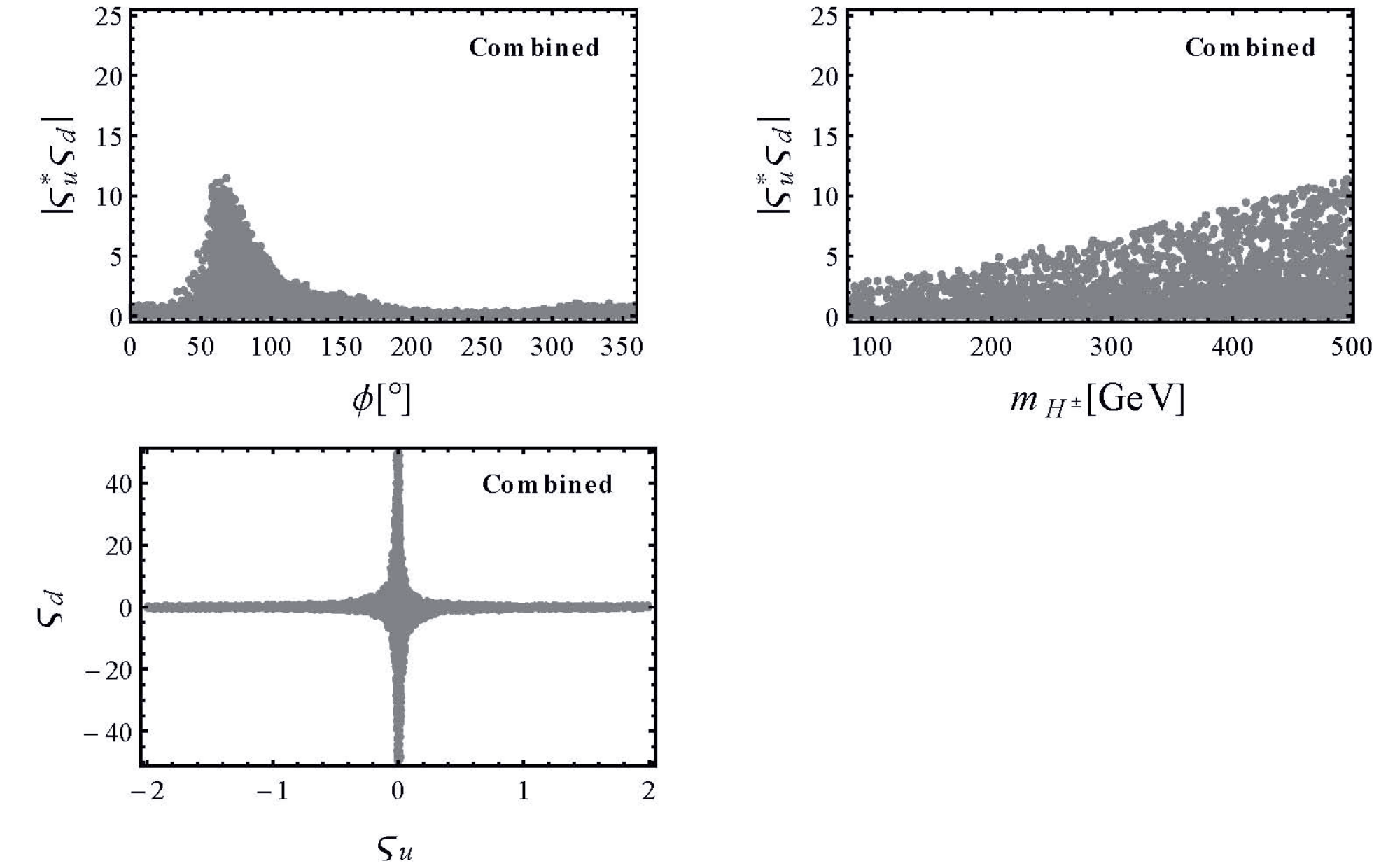}
\caption{\label{fig:final-constraints} \small The combined constraints from $\mathrm{Br}(B\to X_s \gamma)$, $\Delta(K^*\gamma)$ and $\Delta(\rho\gamma)$, see also Fig.~\ref{fig:B2XsG-B2XdG-correlation}.}
\end{figure}

With the current experimental data on the CP-averaged branching ratios and the direct CP and isospin asymmetries as constraints, we have derived bounds on the charged-scalar couplings from these decays. It is found that, while the CP-averaged branching ratios cannot give additional information to the one extracted from inclusive $B\to X_{s,d}\,\gamma$ decays, complementary constraints on the model parameters can be obtained from the isospin asymmetries $\Delta(K^*\gamma)$ and $\Delta(\rho\gamma)$. Thus, we show in Fig.~\ref{fig:final-constraints} the final combined constraints from the three observables $\mathrm{Br}(B\to X_s \gamma)$, $\Delta(K^*\gamma)$ and $\Delta(\rho\gamma)$, where the first two plots correspond to the complex couplings, and the third to real ones. The result confirms explicitly the observation that using the exclusive observables can exclude a significant additional part of the parameter space.

In addition, correlations between the various observables in exclusive $B\to V\gamma$ and inclusive $B\to X_{s,d} \,\gamma$ decays have been investigated, some of which will become relevant with the advent of more precise data.

Experimental progress for these modes is expected from LHCb and the next-generation flavour factories. This will either strengthen the constraints shown here, or will show signs of non-standard  effects. Of special interest in that respect are the two isospin asymmetries, while for the direct CP asymmetries to play a role, also theoretical progress in their calculation would be necessary. The exception is the asymmetry in $B_s\to\phi\gamma$, which is cleanly predicted to be tiny in the SM and provides therefore important information on any NP model with new weak phases in $b\to s$ transitions.

\section*{Acknowledgements}

We thank Roman Zwicky for useful exchanges concerning the hadronic inputs. This work has been supported in part by the Spanish Government~[grants FPA2007-60323, FPA2011-23778 and CSD2007-00042~(Consolider Project CPAN)] and the Generalitat Valenciana [Prometeo/2008/069]. X.~Q.~L. was also supported in part by the National Natural Science Foundation of China~(NSFC) under contract No.~11005032 and the Specialized Research Fund for the Doctoral Program of Higher Education of China~(Grant No.~20104104120001). M.~J. is supported by the Bundesministerium f\"ur Bildung und Forschung~(BMBF). A.~P. acknowledges the support of the Alexander von Humboldt Foundation.

\begin{appendix}

\section*{Appendix: Input parameters}
\label{app:input}

In this appendix, we collect the relevant input parameters for our calculation. We restrict the discussion to the hadronic quantities which dominate the theoretical uncertainties; basic parameters that are known precisely, like meson masses and lifetimes, vector boson masses and gauge couplings, can be found in~\cite{Beringer:1900zz}. We use two-loop running for $\alpha_s$ throughout this paper.

\subsubsection*{CKM matrix elements}

In order to extract values for the CKM parameters $\lambda$, $A$, $\overline{\rho}$, and $\overline{\eta}$, observables insensitive to the additional contributions of the A2HDM have to be used. These include the moduli of CKM matrix elements from super-allowed $\beta$ decays~\cite{Hardy:2008gy} and semileptonic $B$-meson decays with light leptons~\cite{Amhis:2012bh}, and the CP-violating angle $\gamma$ extracted from tree-dominated $B$-meson decays~\cite{Charles:2004jd} as more obvious choices. In addition, we use two loop-induced quantities: the A2HDM contributions to the ratio $\Delta m_d/\Delta m_s$~\cite{Amhis:2012bh} cancel, and also the indirect CP asymmetry in $B_d\to J/\psi K_S$ \cite{Amhis:2012bh} is negligibly affected. A global fit to these constraints yields
\begin{equation}
 A=0.80\pm 0.01\pm 0.01\,, \quad \lambda=0.2254\pm 0.0010\,, \quad
 \overline{\rho}=0.14\pm 0.01\pm 0.03\,, \quad \overline{\eta}=0.336^{+0.014}_{-0.009}\pm 0.010\,.
\end{equation}

\subsubsection*{Heavy quark masses}

For the top-quark mass, the most recent results from the Tevatron and LHC  read~\cite{Aaltonen:2012ra,ICHEP:top}
\begin{equation}
m_t^{\rm Tev}=(173.18\pm 0.56\pm 0.75)~{\rm GeV}\,,\quad m_t^{\rm LHC}=(173.3\pm 0.5\pm 1.3)~{\rm GeV}\,,
\end{equation}
where both numbers are correlated averages of all measurements performed so far. These results still assume the measured value to correspond to the pole mass scheme. For this assumption we add an additional  theoretical uncertainty of $1~{\rm GeV}$. There are ongoing efforts to determine the top mass in a definite scheme; the corresponding uncertainties are not yet competitive, but indicate our moderate increase of the uncertainty~(compared to the larger difference $m_t^{\rm pole}-m_t^{\rm \overline{MS}}$) to be appropriate. We average the two, which yields finally
\begin{equation}
m_t = (173.2\pm0.4\pm1.75)~{\rm GeV}\,.
\end{equation}

For the $b$- and $c$-quark running masses in the ${\rm \overline{MS}}$ scheme, we take the values from~\cite{Beringer:1900zz}
\begin{equation}\label{MSbar-mass}
\overline{m}_{b}(\overline{m}_{b}) = 4.18 \pm 0.03~{\rm GeV}\,, \qquad
\overline{m}_{c}(\overline{m}_{c}) = 1.275 \pm 0.025~{\rm GeV}\,.
\end{equation}
To get the corresponding pole and running quark masses at different scales, we use the NLO ${\rm \overline{MS}}$-on-shell conversion and running formulae collected e.g. in Ref.~\cite{Chetyrkin:2000yt}.

\subsubsection*{Nonperturbative meson parameters}

When discussing exclusive $B\to V \gamma$ decays, the key quantities are the hadronic parameters of the involved mesons, such as heavy-to-light transition form factors and the Gegenbauer moments of their LCDAs, as these constitute major sources of uncertainties and have to be chosen with the corresponding care. Most of these parameters are currently not directly known from experiment and have to be determined by nonperturbative methods like QCD sum rules and lattice QCD.

For the Gegenbauer moments, the relevant values are collected in Table~\ref{tab::gegenbauer}. Note that the values for the $\omega$ meson are not actually calculated, but just assumed to be equal to those of the $\rho$ meson, which is the reason why we doubled the corresponding uncertainties.

\begin{table}[thb]
\begin{center}
\caption{\small Values for the relevant Gegenbauer moments, extracted using QCD sum rules.\label{tab::gegenbauer}}
\vspace{0.2cm}
\doublerulesep 0.8pt \tabcolsep 0.11in
\begin{tabular}{cc c c c c}
\hline\hline
Meson 		& $a_{1,\parallel}(1~{\rm GeV})$& $a_{1,\perp}(1~{\rm GeV})$& $a_{2,\parallel}(1~{\rm GeV})$& $a_{2,\perp}(1~{\rm GeV})$& Ref.\\\hline
$\rho$		& ---							& ---						& $0.15\pm 0.07$				& $0.14\pm 0.06$			& \cite{Ball:1996tb,Ball:2006nr}\\
$K^\ast$	& $0.03\pm 0.02$				& $0.04\pm 0.03$			& $0.11\pm 0.09$				& $0.10\pm 0.08$			& \cite{Ball:2003sc,Ball:2005vx,Ball:2006fz}\\
$\phi$		& ---							& ---						& $0.18\pm 0.08$				& $0.14\pm 0.07$			& \cite{Ball:2007rt}\\
$\omega$	& ---							& ---						& $0.15\pm 0.14$				& $0.14\pm 0.12$			& see text\\
\hline\hline
\end{tabular}
\end{center}
\end{table}

For the heavy-to-light tensor form factors, we use~\cite{Ball:2004rg}
\begin{equation}\label{eq::Foff}
F_{B_q\to V} = \hat{f}_V \left[F_{B_q\to V}^L+\hat{a}_1^L F_{B_q\to V}^{L,a_1}+\frac{\hat{f}_V^\perp}{\hat{f}_V} (F_{B_q\to V}^T+\hat{a}_1F_{B_q\to V}^{T,a_1})\right]\,,\nonumber
\end{equation}
which expresses the results of the sum rule calculation in terms of the corresponding decay constants and Gegenbauer moments~(normalized to their central values in that calculation), making the inclusion of updated evaluations of these quantities possible. The coefficients, including the remaining uncertainties from the sum rule calculation, can be found in the same reference.

We extract the longitudinal decay constants of the vector mesons, $f_V^\parallel\equiv f_V$, from data. For the charged mesons, we can use
\begin{equation}\label{eq::brtau}
\mathrm{Br}(\tau^- \to V^-\nu_\tau) = \tau_\tau \frac{m_\tau^3}{16\pi}\left|V_{UD}\right|^2 f_{V,\parallel}^2 \left(1-\frac{m_V^2}{m_\tau^2}\right)^2\left(1+2\frac{m_V^2}{m_\tau^2}\right)\,.
\end{equation}
Note that this formula holds only in the small width approximation. Regarding the ``branching ratios" to vector mesons, for $K^*$ exists a dedicated analysis~\cite{Jamin:2008qg}, while for $\rho$ we somewhat naively use $\mathrm{Br}(\tau\to\rho\nu)\simeq \mathrm{Br}(\tau\to \pi\pi\nu)-\mathrm{Br}(\tau\to\pi\pi\nu)_{\rm non-res.}$, using the data from \cite{Beringer:1900zz}.

For the neutral vector mesons, the decay constants in question are related to the corresponding radiative decays. We follow in this extraction mainly \cite{Ball:2006eu}.\footnote{Our value for the extracted decay constants differs from Ref.~\cite{Ball:2006eu} as the authors of that paper have got a numerical error in the evaluation~\cite{romanzwicky}.} However, as we do not consider the vector meson mixing in the QCDF analysis, we extract the corresponding decay constants with and without taking this mixing into account, using the difference as an additional theoretical uncertainty. As the analysis is in addition performed in the isospin limit, we do the same for the differences between the decay constants of the charged and neutral $\rho$ mesons.

The transverse decay constants are not accessible experimentally; they are again typically calculated in the frameworks of QCD sum rules or lattice QCD. Generally the calculations for the ratio with respect to the longitudinal decay constant are more stable, therefore we use only the ratios from theory calculations. Note that they are scale dependent, and we take into account the (LL) running via $f_\perp(\mu) = f_\perp(\mu_0) \,\left(\alpha_s(\mu)/\alpha_s(\mu_0)\right)^{4/23}$.

For the $\rho$ and $K^*$ mesons, we use the QCD sum rule results from \cite{Ball:2006eu}. The $\phi$ meson is problematic in these calculations, and we prefer to use the lattice result from \cite{Allton:2008pn}. However, given the relatively large spread with respect to the calculations in \cite{Jansen:2009hr,Becirevic:2003pn,Braun:2003jg}, we increase the theory error to $\sim3\%$. The $\omega$ meson is a special case, as there exists neither a QCD sum rule nor a lattice QCD calculation for its decay constant. It is commonly~(but somewhat crudely) estimated to be equal to the one of the $\rho$ meson. We use this estimate as well, but increase the corresponding uncertainty to account for this.

In Table~\ref{tab::formfactors} we collect the extracted decay constants, as well as the resulting values for the transition form factors. In addition to the values given there, we use the ratio $f_{B_s}/f_{B_d}=1.198\pm0.009\pm0.025$~\cite{Albertus:2010nm,Bazavov:2011aa,Na:2012kp} to determine the decay constant $f_{B_{u,d}}$.

\begin{table}[thb]
\begin{center}
\caption{\label{tab::formfactors} \small Decay constants and form factors for the relevant decays, for details on the extraction, see text. The tensor form factors are given for $B_{u,d} \to \rho,\omega$, but for $B_s\to \phi$.}
\vspace{0.2cm}
\doublerulesep 0.8pt \tabcolsep 0.07in
\begin{tabular}{cc c c c}
\hline\hline
Meson 		& $f_\parallel/{\rm GeV}$ 	& $f_\perp(2~{\rm GeV})/f_\parallel$	& $T_1(0)$					 & Ref.\\\hline
$B_{s}$	& $0.228\pm0.001\pm0.006$	& ---									& ---						& \cite{Bazavov:2011aa,McNeile:2011ng,Na:2012kp}\\
$\rho$		& $0.215\pm0.001\pm0.006$	& $0.70\pm0.04_{\rm th}$				& $0.276\pm0.001\pm0.039$	 & \cite{Beringer:1900zz,Ball:2006eu}\\
$K^\ast$	& $0.209\pm0.007_{\rm stat}$& $0.73\pm0.04_{\rm th}$				& $0.302\pm0.010\pm0.051\,(B_{u,d})$& \cite{Beringer:1900zz,Ball:2006eu}\\
			&							&										& $0.274\pm0.009\pm 0.044\,(B_s)$& \\
$\phi$		& $0.229\pm0.002\pm0.002$	& $0.750\pm0.004\pm0.020$				& $0.335\pm0.003\pm0.043$	 & \cite{Beringer:1900zz,Allton:2008pn}\\
$\omega$	& $0.188\pm0.002\pm0.010$	& $0.70\pm0.10_{\rm th}$				& $0.237\pm0.003\pm0.055$	 & \cite{Beringer:1900zz}$^\dagger$\\
\hline\hline
\end{tabular}
\footnotesize \parbox[thb]{15.5cm} {\vspace{0.15cm} $^\dagger$ See text.}
\end{center}
\end{table}

\end{appendix}

\end{document}